\documentclass[fleqn,usenatbib]{mnras}
\usepackage[T1]{fontenc}
\usepackage{commath}
\usepackage{amsmath}	
\usepackage{amssymb}	
\usepackage{ae,aecompl}
\usepackage{graphicx}	
\usepackage[usenames,dvipsnames,svgnames,table]{xcolor}
\usepackage[table]{xcolor}
\usepackage{verbatim}
\usepackage[font = small, labelfont=bf, labelsep=period]{caption}
\usepackage{subcaption}
\usepackage{breqn}
\usepackage{float}
\usepackage{dblfloatfix}

\usepackage[utf8]{inputenc}
\usepackage{multicol}
\usepackage{bibspacing}
\usepackage{lipsum}
\usepackage{hyperref}
\usepackage{diagbox}
\usepackage[inline]{enumitem}
\usepackage{enumitem}
\usepackage[normalem]{ulem}
\setlist{
	listparindent=\parindent,
	parsep=0pt,
}
\hypersetup{ %
			colorlinks=true,%
			pdfauthor={Nomen Tuum},%
			pdftitle={Accretion onto eccentric binaries},%
			citecolor=blue%
			}

\def\red#1 {\textcolor{red}{#1}\ }  
\def \green#1 {\textcolor{green}{#1}\ }  

\title[Accretion Disks around Eccentric Binary Systems]{Preferential Accretion and Circumbinary Disk Precession in Eccentric Binary Systems}
\author[Magdalena Siwek et al.]{
Magdalena Siwek,$^{1}$\thanks{E-mail: magdalena.siwek@cfa.harvard.edu}
Rainer Weinberger,$^{1,2}$
Diego J. Mu\~{n}oz,$^{3,4,5}$
Lars Hernquist$^{1}$
\\
$^{1}$Center for Astrophysics, Harvard University, Cambridge, MA 02138, USA \\
$^{2}$ Canadian Institute for Theoretical Astrophysics, 60 St. George St., Toronto, ON M5S 3H8, Canada \\
$^{3}$Center for Interdisciplinary Exploration and Research in Astrophysics (CIERA), Northwestern University, Evanston, IL 60208, USA\\
$^{4}$ Facultad de Ingenier\'ia y Ciencias, Universidad Adolfo Ib\'a\~nez, Av.\ Diagonal las Torres 2640, Pe\~nalol\'en, Santiago, Chile\\
$^{5}$ Millennium Institute for Astrophysics, Chile
}

\date{Accepted XXX. Received YYY; in original form ZZZ}

\pubyear{2022}

\begin{document}
\label{firstpage}
\pagerange{\pageref{firstpage}--\pageref{lastpage}}
\maketitle

\begin{abstract}
We present a suite of high resolution hydrodynamic simulations of binaries immersed in circumbinary accretion disks (CBDs). For the first time, we investigate the preferential accretion rate as a function of both eccentricity $e_{\rm b}$ and mass ratio $q_{\rm b}$ in a densely sampled parameter space, finding that when compared with circular binaries, \begin{enumerate*}[label=(\roman*)]
	\item mass ratios grow more efficiently in binaries on moderately eccentric orbits ($0.0 \lesssim e_{\rm b} \lesssim 0.4$),  and
	\item high eccentricities ($e_{\rm b} \gtrsim 0.6 $) suppress mass ratio growth.
\end{enumerate*}
We suggest that this non-monotonic preferential accretion behaviour may produce an observable shift in the mass ratio distributions of stellar binaries and massive black hole binaries.
We further find that the response of a CBD can be divided into three regimes, depending on eccentricity and mass ratio: (i) CBDs around circular binaries always precess freely, whereas CBDs around eccentric binaries 
either (ii) undergo
forced precession or (iii) remain  locked at an angle with respect to the binary periapsis. Forced precession in eccentric binaries is associated with strong modulation of individual accretion rates on the precession timescale, a potentially observable signature in accreting binaries with short orbital periods. We provide CBD locking angles and precession rates as a function of $e_{\rm b}$ and $q_{\rm b}$ for our simulation suite.
\end{abstract}


\section{Introduction}
Numerical simulations of circumbinary disks (CBDs) have consistently shown that the secondary is favoured in the accretion flow, pushing the mass ratios of binaries towards unity (\citealt{Artymowicz1983, Bate2002, Farris2014, Gerosa2015, Munoz2020, Duffell2020}). This accretion behaviour is expected to affect the population statistics of long-lived accreting binaries, such as supermassive black hole binaries (MBHBs). Whether the preferential accretion on the secondary is significant enough to measurably affect the mass ratio distributions of accreting binary populations was unclear, until recently. To quantify the impact of preferential accretion models on binary populations within their lifetimes, \cite{Siwek2020} evolved populations of accreting MBHBs from cosmological simulations (see also \citealt{Kelley2019, Bortolas2021}). \cite{Siwek2020} found that preferential accretion onto the secondary significantly increases the mass ratios of MBHBs prior to merger, affecting predicted amplitudes of the gravitational wave background (GWB, \citealt{Rajagopal1995, Phinney2001, Wyithe2003}) and electromagnetic wave (EM) observations of inspiraling binaries.

However, these results were based on preferential accretion models derived from circular binaries only, as simulations so far have focused on either equal mass binaries on eccentric orbits (e.g. \citealt{Munoz2019b}), or circular binaries with varying mass ratios (e.g. \citealt{DOrazio2013,Farris2014, Munoz2020, Duffell2020}; note however that \citealt{DOrazio2013} did not report preferential accretion rates). A few exceptions exist where small samples of non equal mass, eccentric binaries  were studied \citep{ArtymowiczLubow1996,Roedig2011, Dunhill2015}. In particular, \cite{Dunhill2015} found tantalizing evidence that preferential accretion onto the secondary in an eccentric, unequal mass ratio binary was mildly suppressed, and suggested that this could impact the mass ratio growth in eccentric binary systems. However, no large parameter studies over mass ratio and eccentricity are available, and as a result it is currently not possible to predict the effect of binary accretion on populations of MBHBs, which typically have varying mass ratios and eccentricites. In the following work we investigate the preferential accretion rate of a binary as a function of eccentricity and mass ratio.

We also investigate how the central binary affects the long-term dynamics of the surrounding disk. Simulations have shown that gas eccentricity grows and saturates quickly in CBDs, to remain at steady, significant values  ($e_{\rm d} \sim 0.3$) throughout the lifetime of the disk \citep{Papaloizou2001,MacFadyen2008,Cuadra2009,DOrazio2016, Miranda2016, Thun2017, MunozLithwick2020}. Subsequently, CBDs around circular binaries are found to precess uniformly in simulations (e.g. \citealt{Thun2017, Miranda2016}), at timescales close to the binary quadrupole frequency \citep{MacFadyen2008, MunozLithwick2020}. The disk response appears to change when the central binary is eccentric: \cite{Miranda2016} found that CBDs can apsidally align with equal mass, eccentric binaries. In simulations where CBDs precess around an eccentric binary, periodic accretion variability with orders of magnitude changes in accretion rates are observed (e.g. \citealt{Dunhill2015, Munoz2016}), giving tantalizing prospects for transient observations of MBHBs and stellar binaries. However, these promising results remain restricted to equal mass eccentric binaries. 

Here we study the effect of varying binary eccentricity and mass ratio on the accretion behaviour of the binary and the CBD eccentricity and precession. We present a new suite of high resolution simulations illustrating the long term accretion behaviour of binaries  with varying mass ratios and eccentricities, and evolution of the surrounding CBDs, evolved over thousands of binary orbits. 

This paper is structured as follows. In Section \ref{sec:numerical} we present the initial conditions and numerical methods used to evolve our hydrodynamic simulations. In Section \ref{sec:results} we present results from our simulations, including: the preferential accretion of binaries as a function of mass ratio and eccentricity (Section \ref{sec:prefacc}), associated timescales of mass ratio growth (Section \ref{sec:timescales}), and new findings regarding the locking and forced precession of CBDs around eccentric binaries (Section \ref{sec:precession}).
In Section \ref{sec:discussion} we discuss our results in a broader astronomical context, outlining potential applications and highlighting limitations and caveats to our work.

\section{Numerical Methods}
\label{sec:numerical}
We carry out hydrodynamic simulations of binaries immersed in CBDs using the moving mesh code Arepo \citep{Springel2010, Pakmor2016} in its Navier Stokes version \citep{Munoz2013}, which employs a Voronoi tessellation to generate a grid around a set of discrete mesh-generating points.

The binary is represented by two sink particles with a mass ratio $q_{\rm b}$ and sink radii $R_{\rm s} = 0.03 a_{\rm b}$, moving on a fixed Keplerian orbit with arbitrary eccentricity $e_{\rm b}$. The fraction of gas accreted by the binary and removed from cells within the sink region is a function of the radial separation $R_{\rm ji}$ between the j-th sink particle and i-th gas cell. We define the dimensionless parameter $\gamma = \gamma_0 \times (1 - R_{\rm ji}/R_{\rm s} )^2$, where $\gamma_0 = 0.5$, which determines the fraction of gas removed from each cell within the sink radius at each timestep. 

Our initial accretion disk is a finite, locally isothermal disk in a computational box with open boundary conditions.
Our initial disk surface density $\Sigma (R)$ is similar to that used in \cite{Munoz2020}, 

\begin{equation}
\Sigma (R) = f_{\rm t}  \, \Sigma_0 \Big(\frac{R}{a_{\rm b}}\Big)^{-1/2}  \Big[ 1 - 0.7\, \sqrt{\frac{R}{a_{\rm b}}} \Big].
\label{eqn:rho_initial}
\end{equation}
$f_{\rm t}$ is a tapering function that initializes a finite disk with an inner cavity at a radius $R_{\rm cav} = 2a_{\rm b}$, and an outer edge $R_{\rm outer} \sim 50a_{\rm b}$.  
This profile is chosen so that $\Sigma(R) \propto R^{-1/2} \propto \dot{M}_{\rm b}/\nu(R)$, and matches the steady state solution of an accretion disk at large radii from the binary $R/a_{\rm b} \gg 1$. While the surface density morphology of the inner region of a CBD is expected to depend on the binary parameters, the effect on the outer disk at $R/a_{\rm b} \gg 1$ should be negligible. We therefore use the same surface density profile in equation \ref{eqn:rho_initial} throughout the parameter space we explore. 

Our 2D simulations are evolved in a computational box of size $300a_{\rm b} \times 300a_{\rm b}$, leaving enough space for the disk to viscously spread outward. 
We model the radial and azimuthal initial velocity in the disk similarly to \cite{Munoz2019b}.
The initial azimuthal velocity profile of the disk is Keplerian, with corrections accounting for the quadrupole potential of the binary and the pressure gradient in the disk,

\begin{equation}
v^2_{\Phi}(R) = \frac{G M_{\rm b}}{R}\left[1+3\frac{Q}{R^2}\right] - c^2_s(R)\left[1-\frac{R}{\Sigma}\frac{d\Sigma}{dR}\right],
\label{eqn:v_azimuthal_corrected}
\end{equation}
where $G=1$ is the gravitational constant, $M_{\rm b} = 1$ is the mass of the binary, $c_{\rm s}$ the sound speed and $Q$ accounts for the quadrupole moment of the binary and is given by

\begin{equation}
Q = \frac{a_{\rm b}^2}{4} \frac{q_{\rm b}}{(1+q_{\rm b})^2} \Big(1 + \frac{3}{2}e_{\rm b}^2\Big).
\label{eqn:v_quadrupole_correction}
\end{equation}

The initial radial velocity profile accounts for the viscous drift in the disk, and is given as follows

\begin{equation}
v_R(R) = \frac{1}{R\Sigma} \frac{\partial}{\partial R} \Big( \nu \Sigma R^3 \frac{d\Omega}{dR}\Big) \Big[\frac{d}{dR} (R^2\Omega)\Big]^{-1},
\label{eqn:v_viscous_drift}
\end{equation}
where $\nu$ is the kinematic viscosity coefficient and $\Omega$ the Keplerian frequency.

We choose a locally isothermal equation of state
\begin{equation}
	c_{\rm s}^2(R) = h^2 \, \abs{\phi_{\rm b}(R)},
\end{equation}
where $h = 0.1$ is the disk aspect ratio, and $\phi_{\rm b}$ the binary potential given by the expression

\begin{equation}
\phi_{\rm b}(R) = -\frac{GM_1}{R_1} - \frac{GM_2}{R_2},
\label{eqn:binary_potential}
\end{equation}
where $R_1$ and $R_2$ are the distances to the primary (higher mass) component of mass $M_1$, and secondary (lower mass) component of mass $M_2$, respectively. $M_{\rm b}$ is the total mass of the binary, which we set to unity so that $M_{\rm b} = M_1 + M_2 \equiv 1$. We set the binary semi-major axis $a_b = 1$, fixing the period of the binary to $P_b = 2\pi$. The softening length is $R_{\rm soft} = 0.025$, smaller than the sink radius $R_{\rm s} = 0.03$, and we use a smooth spline function to soften the potential, such that the the gravitational force becomes exactly Newtonian when $R > R_{\rm soft}$ \citep{Springel2001}, ensuring accuracy of the potential even during pericenter approach of highly eccentric binaries.

The viscosity is modeled with an $\alpha$-prescription both in the CBD around the binary and on circumsingle disk (CSD) scales with a smooth transition in between. We set $\alpha = 0.1$ in all our simulations.  More details on the viscosity implementation can be found in appendix section \ref{sec:appendix_viscosity}.

We further choose the mass resolution in the central region of our simulations as small as $m_{\rm res} =5 \times 10^{-6} \Sigma_0 \, a_{\rm b}^2$. With typical circumsingle disk (CSD) densities reaching $\Sigma \gtrsim  5 \Sigma_0 $ at pericenter approach (compare with surface densities at apocenter, Figure \ref{fig:all_sims_snap}), we follow the accretion flow near the binary down to scales $r_{\rm cell} \lesssim 1 \times 10^{-3} a_{\rm b}$, thus well resolving the CSDs around each sink particle.

\subsection{Steady State}
\label{sec:steady}
The viscous time at a distance $R$ from the binary barycenter is defined as follows,

\begin{equation}
t_{\nu}(R) = \frac{4}{9}\frac{R^2}{\nu(R)}.
\label{eqn:viscous_time}
\end{equation}

Given our choice of constants, and for an integration time of  up to  $t_{\rm sim} = 10000 P_{\rm b}$  in our simulations, our disks are viscously relaxed out to a radius up to $R_{\nu} \sim 27 a_{\rm b}$, well into a regime where the disk can be considered axisymmetric. Since our disks are finite and thus viscously spreading, the system never reaches a true steady state. However, a quasi-steady state is reached when the total accretion rate $\dot{M}_{\rm b}$ of the binary immersed in a disk of initial mass $M_{\rm d,0}$ (within the characteristic radius $R_{\rm d,0}$) follows the time dependence of a viscously spreading disk (e.g.  \citealt{Hartmann1998, Andrews2009}),
\begin{equation}
\label{eqn:visc_spread_mdot}
	\dot{M}_{0, \rm visc} = \frac{3(2-\beta)\nu_0}{2} \Big( \frac{M_{\rm d,0}}{R^2_{\rm d,0}}\Big) \Big( 1 + \frac{t}{t_{\nu, 0}} \Big)^{\frac{-5+2\beta}{4 - 2\beta}}
\end{equation}

\begin{figure} 
	\centering
	\includegraphics[width=1.0\columnwidth]{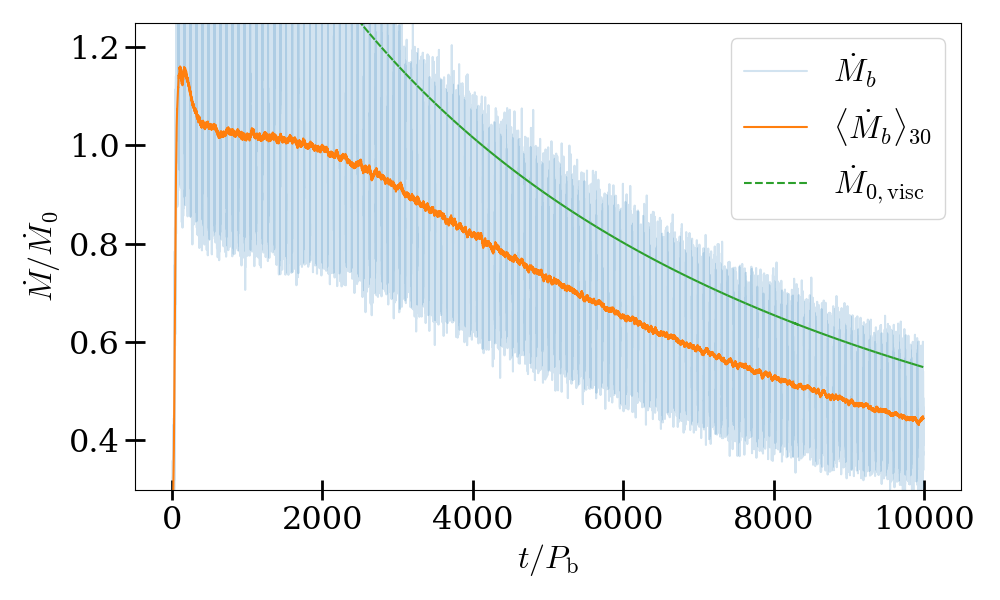}
	\caption{We show the raw accretion rates (grey line) from our simulation of a circular, equal mass ratio binary alongside the moving mean over $30 P_b$ (orange line), and the slope of the accretion rate evolution from equation (\ref{eqn:visc_spread_mdot}) as a green dashed line with $\beta = 0.8$. The long term decay is due to the viscously spreading outer boundary of the disk and matches the analytical model after an initial phase. }
	\label{fig:visc_spreading_acc}
\end{figure}

In Figure \ref{fig:visc_spreading_acc} we present the long term accretion rate of a $e_{\rm b} = 0.00$, $q_{\rm b} = 1.00$ binary compared with the theoretical estimate from equation (\ref{eqn:visc_spread_mdot}). After an initial phase, the accretion rate measured in our simulations shows good agreement with the slope in equation \ref{eqn:visc_spread_mdot}, and thus matches the analytic model of a viscously spreading accretion disk. 

\section{Results}
\label{sec:results}

\subsection{Literature Comparison of Preferential Accretion in Circular Binaries}
\label{sec:literature}
In this work we present a new suite of 50 hydrodynamic simulations, studying the ratio of secondary versus primary accretion rates (preferential accretion) as a densely sampled function of mass ratio and eccentricity.
\begin{figure} 
	\centering
	\includegraphics[width=1.0\columnwidth]{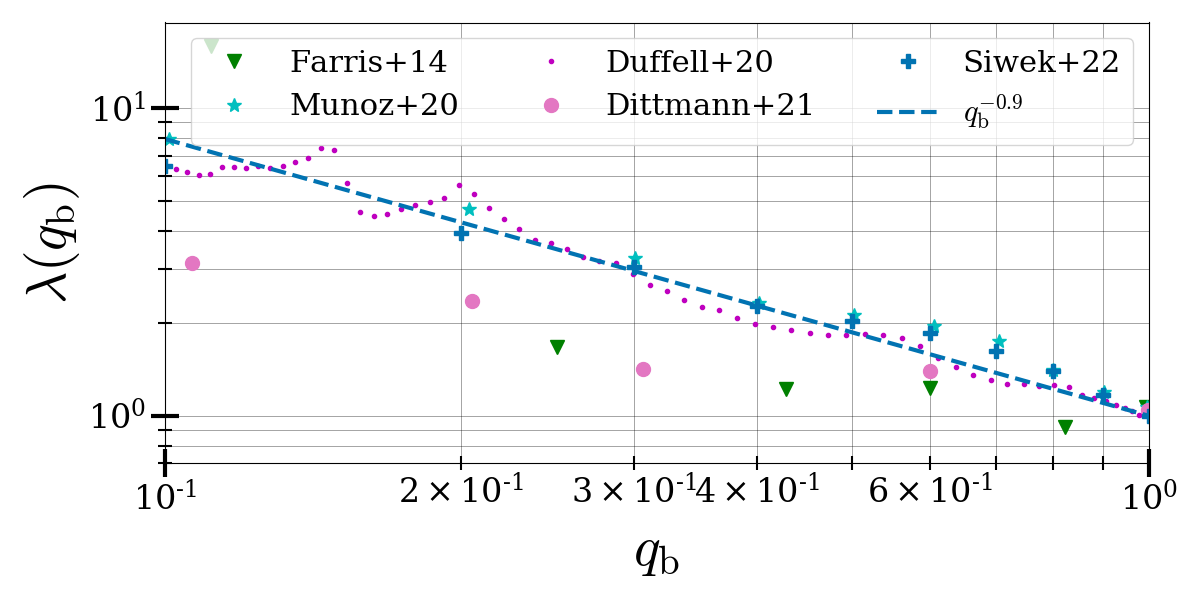}
	\caption{Literature comparison of preferential accretion rates as a function of mass ratio for circular binaries. This work (Siwek+22 in the legend) finds excellent agreement with similar work by \protect\cite{Munoz2020} (light blue stars) and \protect\cite{Duffell2020} (magenta dots), and is well approximated by a fitting function with $\lambda_{\rm fit}(q_{\rm b}) = q_{\rm b}^{-0.9}$ (blue dashed line). These most recent works measure systematically higher values of $\lambda$ compared with \protect\cite{Farris2014}, and \protect\cite{Dittmann2021}.}
	\label{fig:literature_and_this_work_pref_acc}
\end{figure}
\begin{figure*}
	\centering
	\includegraphics[width=1.0\textwidth]{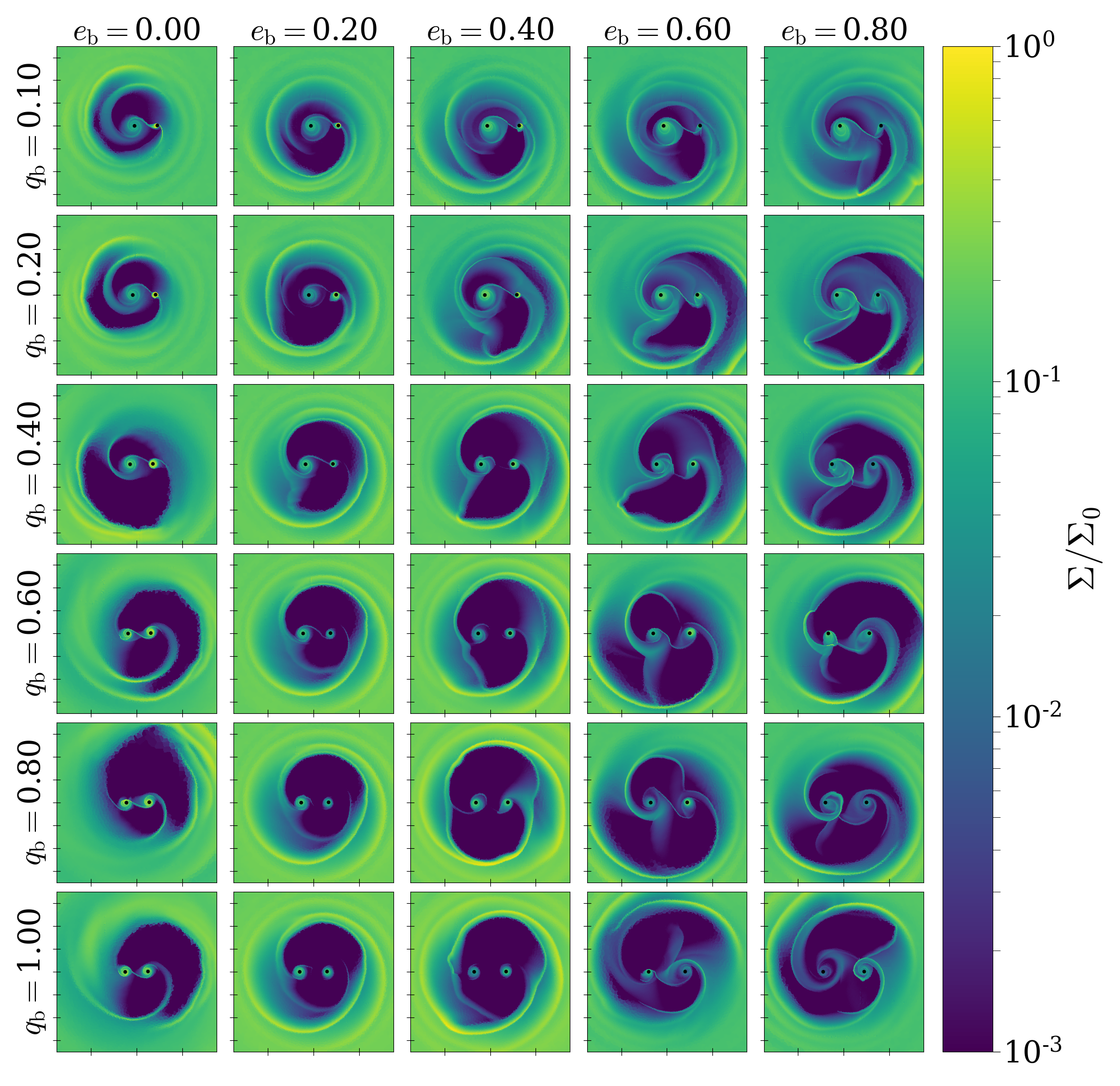}
	\caption{Surface density projection plots from our suite of hydrodynamic simulations, showing binaries with eccentricities $e_{\rm b}: [0.0 \rightarrow 0.8]$ increasing along the horizontal axis,  and mass ratios $q_{\rm b}: [0.1, 0.2, 0.4, 0.6, 0.8, 1.0]$, increasing vertically downwards. All binaries shown have been evolved for 2000 $P_b$, and are pictured here at apocenter. For simulations of a given eccentricity, the cavity around the binary tends to have lower surface density and more compact CSDs the higher the mass ratio. Comparing simulations of constant mass ratio, increasing eccentricity increases the extent of the cavity, but forms more extended CSDs and streams around the binary.}
	\label{fig:all_sims_snap}
\end{figure*}
Before presenting our eccentricity parameter study, we first compare to the existing literature in the case of circular binaries.

In Figure \ref{fig:literature_and_this_work_pref_acc} we give an overview of recent preferential accretion studies for circular binaries. We define the preferential accretion rate as $\lambda(q_{\rm b}) \equiv \dot{M}_2/\dot{M}_1$.
We compare our work (blue crosses, labeled here Siwek+22) with previous work by \cite{Farris2014} (green triangles), \cite{Munoz2020} (blue stars), \cite{Duffell2020} (magenta dots) and \cite{Dittmann2021} (pink circles). Our work yields excellent agreement with \cite{Munoz2020} and \cite{Duffell2020}, and is well approximated by a fitting function $\lambda_{\rm fit}(q_{\rm b}) = q_{\rm b}^{-0.9}$ (blue dashed line).
Our work, as well as the results from \cite{Munoz2020} and \cite{Duffell2020}, differ slightly from \cite{Farris2014} and \cite{Dittmann2021}. 

All simulations shown in figure \ref{fig:literature_and_this_work_pref_acc} use an aspect ratio $h = 0.1$, a locally isothermal equation of state and $\alpha = 0.1$. We briefly discuss some of the differences between our simulations and \cite{Farris2014} and \cite{Dittmann2021}, including resolution, length of simulation and viscosity treatment. 
 \begin{enumerate*}[label=(\roman*)]
	\item Resolution: Our simulations resolve the gas cells in the cavity region down to scales $\lesssim 10^{-3} a_{\rm b}$,  while both \cite{Farris2014} and \cite{Dittmann2021} resolve the CSDs to at most $5 \times 10^{-3} a_{\rm b}$ (in all mass ratios greater than $q_{\rm b} > 0.1$).  
	\item Length of simulation: \cite{Farris2014} evolve their binaries over $500\,P_{\rm b}$, \cite{Dittmann2021} over $2000\,P_{\rm b}$, while our simulations are evolved over $10000\,P_{\rm b}$. 
	Finally, \item viscosity treatment may be the most important difference in the 3 differing simulations: Our simulations model all three disks, including the outer CBD and the inner CSDs, with an $\alpha$-disk profile (see appendix \ref{sec:appendix_viscosity} for more details). \cite{Farris2014} treat only the outer CBD as an $\alpha$-disk, and compute the viscosity in the inner cavity region using the gravitational potential of a stationary point mass at the origin rather than the time-varying potential of a binary. Within the CBD cavity, this might lead to some deviations in the accretion behaviour. \cite{Dittmann2021} use a constant $\nu $ viscosity implementation, however this is similar to the method in \cite{Duffell2020}, whose results agree well with those presented in this work.
\end{enumerate*}

We suspect that a combination of different numerical methods may cause the small differences seen in Figure \ref{fig:literature_and_this_work_pref_acc}. However, the excellent agreement between our work, \cite{Munoz2020} and \cite{Duffell2020}, and the overall agreement across all simulations that secondary accretion is favoured across all mass ratios, is encouraging and suggests that simulations of preferential accretion in circular binaries are quickly converging to a common result.

\subsection{Preferential Accretion in Eccentric Binaries}
\label{sec:prefacc}
We study the accretion flow onto circular and eccentric binaries of varying mass ratios using a suite of 50 high resolution hydrodynamic simulations of binaries embedded in accretion disks, evolved for up to $10000\, P_b$. In Figure \ref{fig:all_sims_snap} we show the surface density plots of a subsection of these simulations at $2000\, P_b$ (well into steady-state in the domain shown). In the left-most column ($e_{\rm b} = 0.0$) we show circular binaries with increasing mass ratios (going from top to bottom). In line with previous simulations (e.g. \citealt{MacFadyen2008}), we observe the formation of an eccentric precessing disk, visible in these snapshots due to the asymmetry in the cavity morphology. As mass ratios increase, the cavity region gets larger and an over-density (or ``lump") develops at its edge, as seen in previous studies (e.g. see the in-depth discussion of the overdensity formation in \citealt{Shi2012}).  We find that for a given mass ratio, the extent of the cavity region increases with eccentricity, however with less compact CSDs and more streams filling the cavity. For binaries of a given eccentricity, increases in mass ratio lead to larger cavities with lower surface density, and more compact CSDs.

\begin{figure*}
	\centering
	\includegraphics[width=1.0\textwidth]{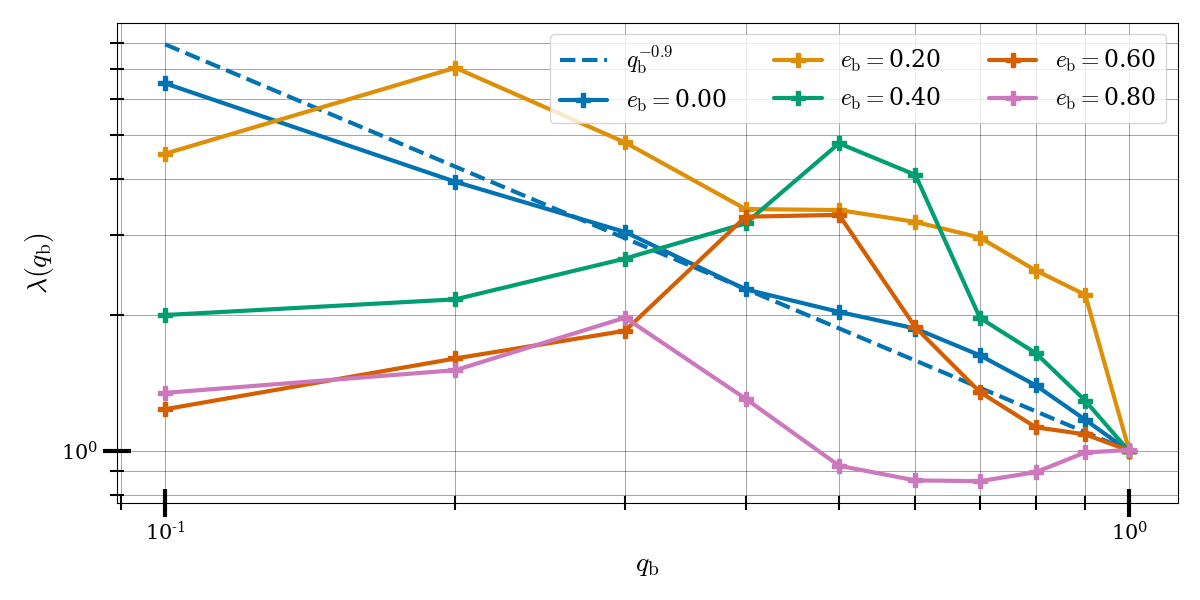}
	\caption{Preferential accretion rates of binaries with eccentricities $e_{\rm b}: [0.0 \rightarrow 0.8]$, as a function of mass ratio. While the circular binary case (blue line) follows the usual fit $\lambda_{\rm fit}(q_{\rm b}) = q_{\rm b}^{-0.9}$ closely, even a low eccentricity in the binary orbit changes this behaviour significantly. For binaries with $e_{\rm b} = 0.2$ (yellow line), the preferential accretion rate is boosted as soon as $q_{\rm b} \gtrsim 0.1$. Moderate eccentricities ($e_{\rm b} =0.4$, $e_{\rm b} = 0.6$; green and orange lines respectively) show a non-monotonic functional form in $\lambda(q_{\rm b})$: the preferential accretion rate exhibits a turnover around $q_{\rm b} \sim 0.3$, below which the preferential accretion of the secondary is decreased compared to binaries on circular orbits. Above $q_{\rm b} \gtrsim 0.3$, the preferential accretion is boosted, and peaks near $q_{\rm b} \sim 0.5$. Once very high eccentricities are reached ($e_{\rm b} = 0.8$, pink line), the preferential accretion rate is lower over the entire mass ratio regime tested here.}
	\label{fig:eccentric_pref_acc}
\end{figure*}

After $\sim 2000\,P_{\rm b}$, we measure the preferential accretion rate $\lambda$ by averaging over another $\sim 500-1000P_{\rm b}$, depending on transient accretion variability associated with CBD precession (see Section \ref{sec:precession}). We plot the result in Figure \ref{fig:eccentric_pref_acc}: the blue dashed line traces the fit to the literature from Figure \ref{fig:literature_and_this_work_pref_acc}, and the blue solid line shows our simulation results for circular binaries, which closely follow the fitting function $\lambda_{\rm fit}(q_{\rm b}) = q_{\rm b}^{-0.9}$. Increasing eccentricity to $e_{\rm b} = 0.2$ (yellow line) changes the preferential accretion of binaries, largely in favour of the secondary: at the lowest mass ratio tested, $\lambda$ decreases relative to the circular binary, but at all other mass ratios, $\lambda$ is significantly increased, implying that the secondary increases its share of the total accretion rate. At both $e_{\rm b} = 0.4$ and $e_{\rm b} = 0.6$ (green and orange lines) the low-$q_{\rm b}$ turnover seen in  the $e_{\rm b} = 0.2$ case shifts to $q_{\rm b} \sim 0.3$, below which $\lambda$ is suppressed. Going to higher mass ratios, the preferential accretion rates increase until they peak near $q_{\rm b} \sim 0.5$, before converging again with the circular preferential accretion curve.

At the highest eccentricity tested ($e_{\rm b} = 0.8$; pink line), $\lambda$ is significantly decreased over the entire mass ratio range shown, and between $0.5 \lesssim q_{\rm b} \lesssim 0.9$, the secondary accretes less than the primary. Despite this, the ratio $\frac{\lambda(q_{\rm b})}{q_{\rm b}} \geq 1 $ everywhere in the mass ratio range tested, and thus the mass ratios of binaries on highly eccentric orbits still increase through accretion. However, since the accretion rate is much more evenly divided between primary and secondary, mass ratios of highly eccentric binaries are expected to grow more slowly when compared with binaries on circular orbits.

\begin{figure} 
	\centering
	\includegraphics[width=1.0\columnwidth]{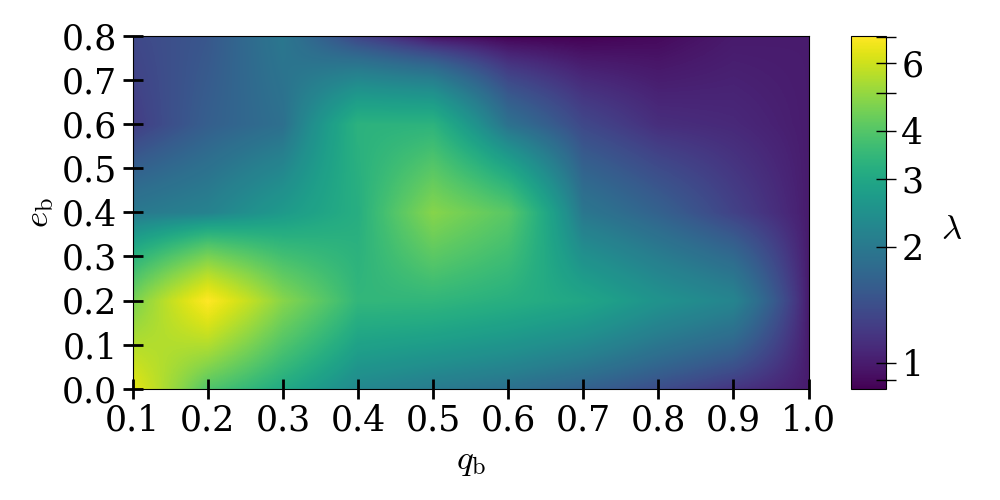}
	\caption{Colormesh of $\lambda$ as a function of $q_{\rm b}$ (horizontal axis) and $e_{\rm b}$ (vertical axis). Preferential accretion is highest in low eccentricity binaries and mass ratios, with a second, lower magnitude peak at moderate mass ratios and eccentricity. At high eccentricity and mass ratios, both primary and secondary accrete at approximately the same rate.}
	\label{fig:interp_lambda}
\end{figure}

While we have tested a fairly densely sampled parameter space in eccentricity and mass ratio, we also interpolate our results so they can be applied to binaries with any orbital parameters. We visualize the 2D interpolation of $\lambda$ in $q_{\rm b}$ and $e_{\rm b}$ in Figure \ref{fig:interp_lambda}. Here we show mass ratio on the horizontal axis, and eccentricity on the vertical axis, and indicated the magnitude of $\lambda$ with a colormap. The `hotspot' lies at the point $q_{\rm b} = 0.2, \, e_{\rm b} = 0.2$, in line with the increased preferential accretion of low eccentricity, low mass ratio binaries in Figure \ref{fig:eccentric_pref_acc}. A second bright spot near $q_{\rm b} = 0.5, \, e_{\rm b} = 0.4$ is associated with the peak in $\lambda$ we find prior to the low-mass ratio turnover in moderate-high eccentricity binaries, also seen in Figure \ref{fig:eccentric_pref_acc}. At the highest eccentricities, preferential accretion is nearly uniformly low, with only a small increase near $q_{\rm b} \sim 0.3$.

Our results show that mass ratio growth through accretion is non-monotonic in eccentricity: $\lambda(q_{\rm b})$ increases at low-moderate eccentricities, and is suppressed in highly eccentric binaries. This may have implications for population properties of accreting binaries occurring on all scales in the Universe. However, we caution that our results are based on our chosen values of disk aspect ratio and viscosity ($h = 0.1$ and $\alpha = 0.1$), and isothermal treatment of the energy equation, while the conditions in stellar or AGN disks may differ from our assumptions.

\subsection{Timescales of mass ratio growth}
\label{sec:timescales}
The mass ratio evolution of a binary with total mass $M_{\rm b}$ depends on the total accretion rate $\dot{M}_{\rm b}$ and preferential accretion factor $\lambda \equiv \frac{\dot{M}_2}{\dot{M}_1}$,

\begin{equation}
\dot{q}_{\rm b}(t) = \frac{(1+q_{\rm b}(t)) (\lambda (q_{\rm b})- q_{\rm b}(t))}{(1 + \lambda(q_{\rm b}))} \frac{\dot{M}_b}{M_b}.
\label{eqn:mass_ratio_evol_acc}
\end{equation}
The mass ratio as a function of time of a binary starting at some initial mass ratio $q_{b,0}$ at a time $t_0$ is then computed by integrating,

\begin{equation}
q_{\rm b}(t)  = q_{\rm b,0} + \int_{t_0}^{t} \dot{q}_{\rm b}(t') dt'.
\label{eqn:mass_ratio_evol}
\end{equation}

\begin{figure*}
	\centering
	\includegraphics[width=1.0\textwidth]{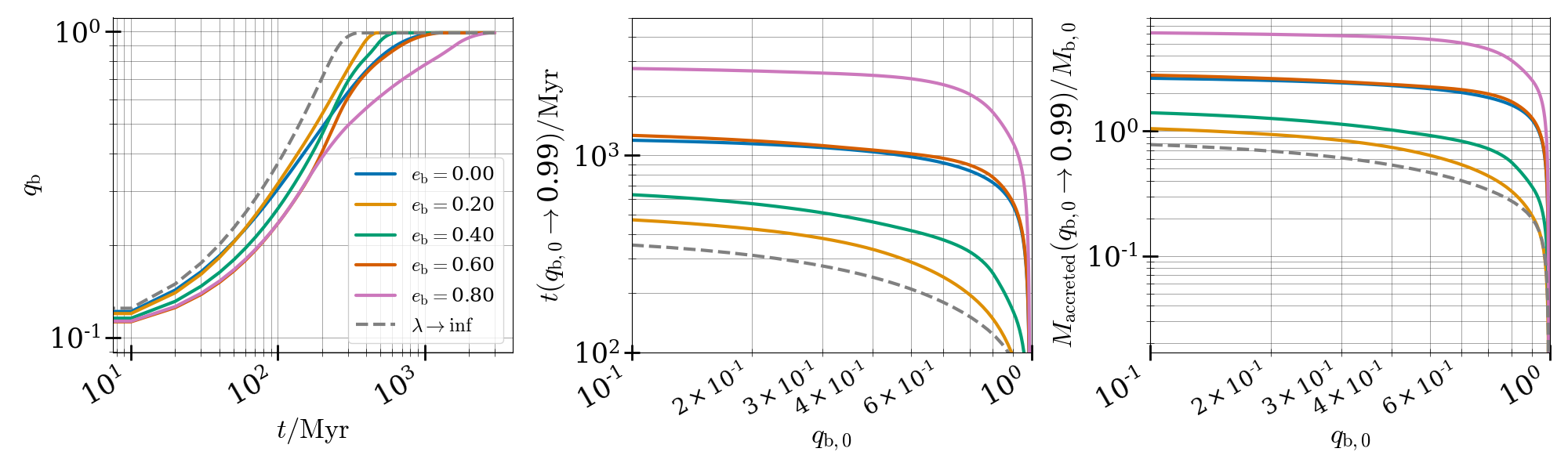}
	\caption{Left panel: Numerical solution to the ODE relating mass ratio evolution to preferential accretion, assuming an accretion rate of 10\% Eddington.
		Middle panel: Timescales for a binary with an initial mass ratio $q_0$ to evolve to a final mass ratio $q_{\rm final} \rightarrow 0.99$, assuming various eccentricities. Binaries with low-moderate eccentricities can equalize their mass ratios $2-3$ times faster than circular binaries. As eccentricity grows beyond $e \simeq0.4$, so does the timescale to mass ratio equalization. Highly eccentric binaries can take up to $2$ times longer to equalize their mass ratios.
		Right panel: Mass accreted by a binary with an initial mass ratio $q_0$ to evolve to a final mass ratio $q_{\rm final} \rightarrow 0.99$. }
	\label{fig:timescale_to_equalizing}
\end{figure*}

Given equation (\ref{eqn:mass_ratio_evol_acc}), the mass ratio evolution as a function of time as shown in Figure \ref{fig:timescale_to_equalizing} (left panel) can be obtained by interpolating $\lambda(q_{\rm b})$ shown in Figure \ref{fig:eccentric_pref_acc}, and incrementally evolving the mass ratio forward in time. In our calculation we assume a fixed total accretion rate at $10 \%$ the Eddington value. This yields a time to mass ratio equalization as low as $\sim 500$ Myr for our lowest eccentricity binary $e_{\rm b} = 0.2$,  and up to 3 Gyr in highly eccentric binaries with $e_{\rm b} = 0.8$. This is further illustrated in the middle panel of Figure \ref{fig:timescale_to_equalizing}: here we integrate the time to mass ratio equalization (vertical axis) as a function of the initial mass ratio $q_{b,0}$ (horizontal axis). We find that eccentric binaries with $e_{\rm b} = 0.2$ equalize their mass ratios up to a factor $\sim 3$ faster compared to circular binaries. As already seen in the left panel, increasing eccentricities further does not decrease the timescales: The shortening of the timescale saturates between $e_{\rm b} = 0.2$ and $e_{\rm b} = 0.4$, after which timescales to mass ratio equalization increase beyond the time given for a circular binary. For high eccentricities ($e_{\rm b} = 0.8$), the timescale to mass ratio equalization is as much as $3\times$ slower compared to a circular binary. Comparing low eccentricity ($e_{\rm b} = 0.2$) to high eccentricity ($e_{\rm b} = 0.8$) mass ratio equalization timescales, we find nearly an order of magnitude difference. 

We also provide the mass that must be accreted by a binary of initial mass ratio $q_{\rm b, 0}$ to equalize its mass ratio (right panel). We find that the $e_{\rm b} = 0.2$ binaries only accrete an amount of gas similar to the total binary mass to equalize the mass ratio of a $q_{\rm b,0} = 0.1$ binary. Even if the entirety of the gas was accreted by the secondary, and $\lambda \rightarrow \infty$, the binary would have to accrete an additional $\sim 80 \%$ of its initial mass in order to reach a mass ratio of unity, as also shown in the gray dashed lines in figure \ref{fig:timescale_to_equalizing}. In the right hand panel, we show that the mass that must be added to the binary in our $e_{\rm b} = 0.2$ simulation to achieve equal mass ratios is only marginally larger than the mass accreted in the $\lambda \rightarrow \infty$ model. This implies that, given a finite gas reservoir, the mass ratio growth rate for a $e_{\rm b} = 0.2$ binary is nearly as efficient as in a scenario where all of the mass is accreted by the secondary.


\subsection{Free Precession, Forced Precession and Locked Disks}
\label{sec:precession}
Similar to previous work (e.g. \citealt{Papaloizou2001,MacFadyen2008, Cuadra2009, Shi2012,Thun2017}), we find that disk eccentricity can grow to significant values ($e_{\rm d} \sim 0.3 - 0.5 $ in the inner region of the CBD), and that the disk precesses around the binary in some, but not all, cases.

\begin{figure*} 
	\centering
	\includegraphics[width=1\textwidth]{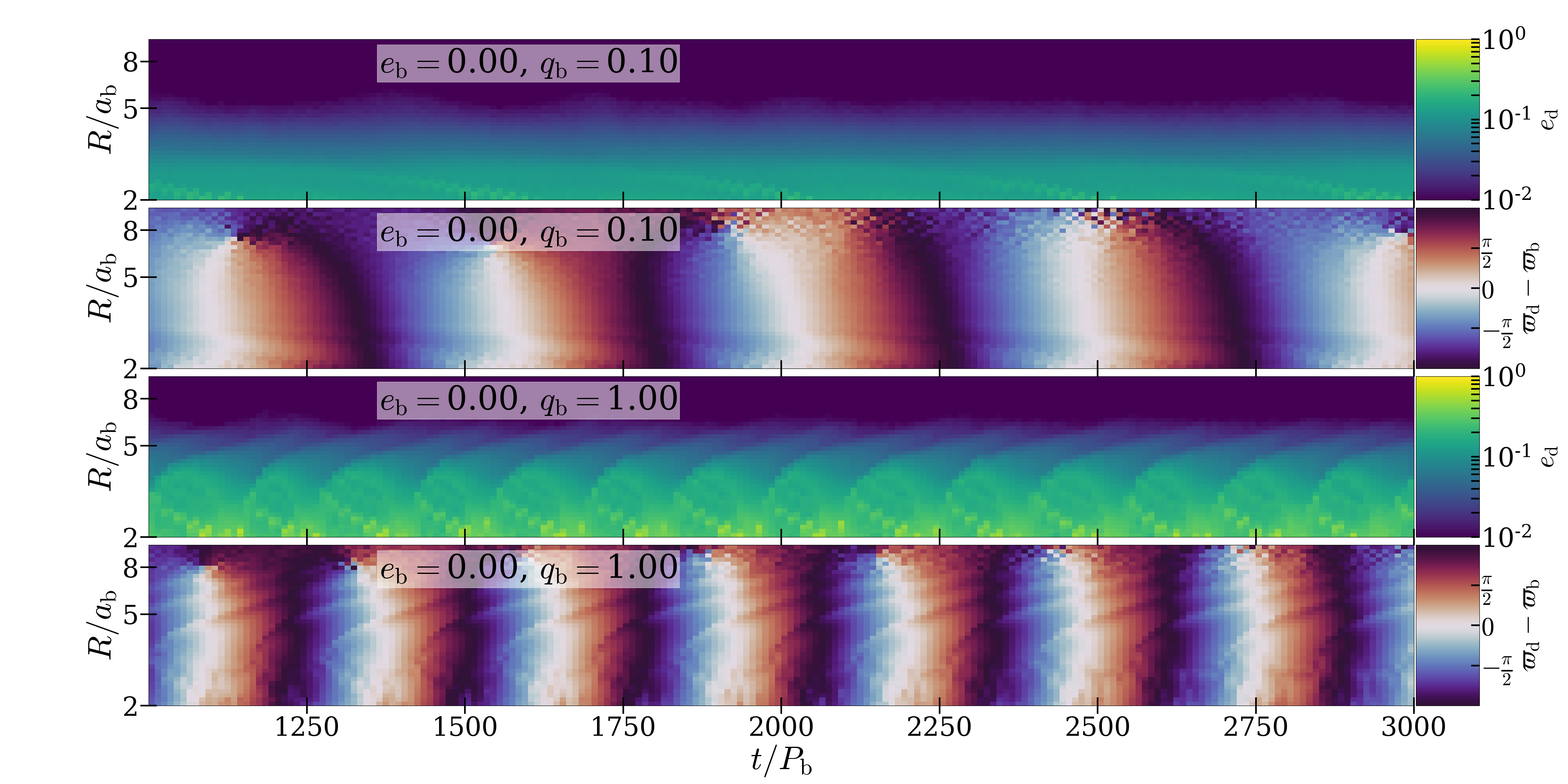}
	\caption{Spacetime diagrams of eccentricity and phase angle , showing free precession of the CBD around circular binaries. Top panel: eccentricity of CBD around low mass ratio binary with $q_{\rm b} = 0.1$. Second from top: Phase angle of CBD around low mass ratio binary with $q_{\rm b} = 0.1$. Third from top: eccentricity of CBD around equal mass ratio binary. Bottom panel: Phase angle of CBD around equal mass binary.}
	\label{fig:disk_precession_circular}
\end{figure*}

We show the eccentricity of CBDs as a space-time diagram in the top and 3rd panel of Figure \ref{fig:disk_precession_circular}, for low mass ratio ($q_{\rm b} = 0.1$; top) and equal mass ratio binaries (3rd from the top) on circular orbits. Similar to \cite{MunozLithwick2020} and \cite{Miranda2016}, we obtain the space-time eccentricity maps by first calculating the Laplace-Runge-Lenz vector (LRL; equivalent to the eccentricity vector) of each gas cell in each snapshot of our simulation:

\begin{equation}
	\textbf{e}_i = \frac{\textbf{v}_i\times(\textbf{r}_i\times\textbf{{v}}_i)}{G M_{\rm b}} -\frac{\textbf{r}_i}{\abs{\textbf{r}_i}}.
	\label{eqn:lrl_vector}
\end{equation}

We bin and average the LRL vectors in radial direction before taking the magnitudes of the resulting vectors. This yields a map of the scalar eccentricity in the disk as a function of time and radius as shown in Figure \ref{fig:disk_precession_circular}. We find that the equal mass ratio binary produces higher eccentricity and eccentricity variability in the disk.

To evaluate the precession of the disks around the circular binaries in Figure \ref{fig:disk_precession_circular}, we study the spatial alignment of the orbits on which the gas in the CBD moves, relative to the binary orbit, as a function of time. Specifically, we calculate the longitude of pericenter of the ellipse on which the gas orbits from the eccentricity vector in equation \ref{eqn:lrl_vector}. The longitude of pericenter of the binary is constant, and set to $\varpi_{\rm b} \equiv \pi$. The difference between the longitude of pericenter of the disk ($\varpi_{\rm d}$)  and that of the binary ($\varpi_{\rm b}$) defines the phase angle $\varpi_{\rm p} \equiv \varpi_{\rm d} - \varpi_{\rm b}$. We plot this quantity as a function of time and radius in the disk, this time shown in panels 2 and 4 of Figure \ref{fig:disk_precession_circular}, for the low- and equal mass ratio binaries respectively.
The resulting space-time diagram of  the phase angle $\varpi_{\rm p}$ shows that CBDs around both low and equal mass ratio circular binaries exhibit prograde, coherent precession of the inner disk over the period of $2000\, P_{\rm b}$ shown, as previously seen in e.g. \cite{Miranda2016} and \cite{MunozLithwick2020}. 

We find that the lower mass ratio binary induces slower CBD precession than the equal mass ratio binary. While the physics that drives CBD precession is not fully understood, it has been found to occur on the timescale of the quadrupole precession frequency evaluated at the inner edge $R_{\rm cav}$ of the CBD \citep{MacFadyen2008}. The lower CBD precession frequency we find around lower binary mass ratio thus follows the expected behaviour from the quadrupole precession frequency (e.g.  \citealt{MoriwakiNakagawa2004}),
\begin{equation}
\label{eqn:quadrupole_frequency}
	\omega_Q = \frac{3}{4} \frac{q_{\rm b}}{(1+q_{\rm b})^2} \Big( 1 + \frac{3}{2} e_{\rm b}^2\Big) \Big(\frac{a_{\rm b}}{R_{\rm cav}}\Big) ^{2}\Omega_{\rm cav},
\end{equation}
which predicts that CBDs should precess faster around binaries with larger mass ratios and eccentricities.

\begin{figure*} 
	\centering
	\includegraphics[width=1\textwidth]{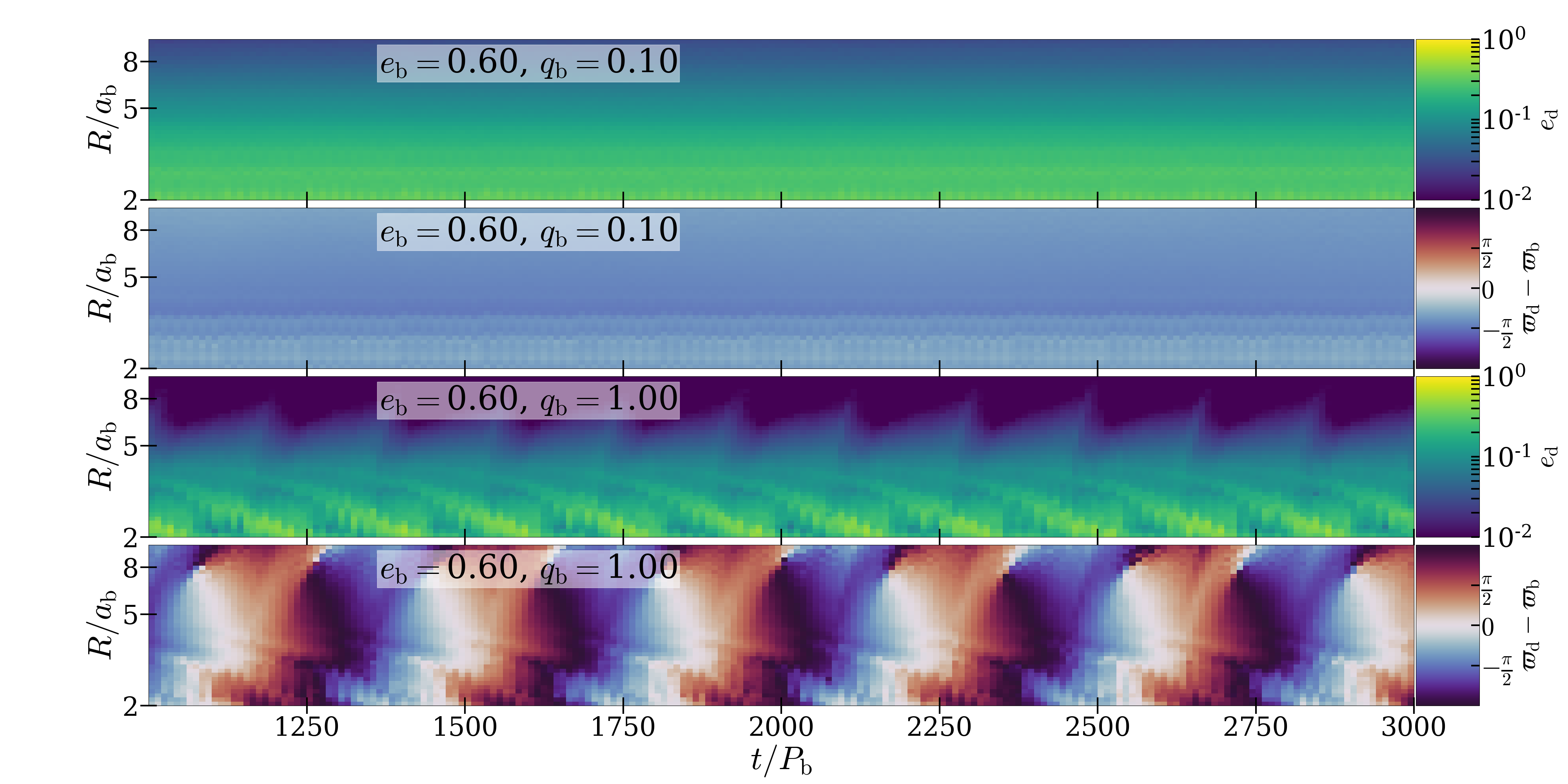}
	\caption{Locking and forced precession of the CBD around binaries with $e_{\rm b} = 0.6$. Top panel: locked CBD around low mass ratio binary with $q_{\rm b} = 0.1$. Bottom panel: forced precession of CBD around equal mass binary, $q_{\rm b} = 1.0$.}
	\label{fig:disk_precession_eccentric}
\end{figure*}

In Figure \ref{fig:disk_precession_eccentric} we show space-time diagrams of CBD eccentricity  around binaries on eccentric orbits. We choose binary eccentricity $e_{\rm b} = 0.6$, and select one low mass ratio (top panel; $q_{\rm b} = 0.1$) and one equal mass ratio (third panel from the top) binary.  When compared with the  low mass ratio, circular binary shown in the top panel of Figure \ref{fig:disk_precession_circular}, the CBD around the low mass ratio eccentric binary has higher eccentricity. Also contrary to the circular case, we find no eccentricity fluctuation in time -- the disk eccentricity is completely steady for thousands of orbits.

To quantify the effect of binary eccentricity on CBD eccentricity, we show the radial profiles in the low mass ratio ($q_{\rm b} = 0.1$) case for two binary eccentricities, $e_{\rm b}= 0$ and $e_{\rm b} = 0.6$, in Figure \ref{fig:e_vs_r}. The inner region of the CBD around the circular binary agrees with the exponential eccentricity profile derived in \cite{MunozLithwick2020}, $e_{\rm d} \propto R^{-5/4} \exp{-(R/\lambda_{\rm d})^{3/4}}$. $\lambda_{\rm d}$ represents the inner tapering radius of the power law, and is dependent upon the ratio of the characteristic frequencies, $\omega_{\mathcal{P}}/\omega_{\mathcal{Q}}$, where $\omega_{\mathcal{P}}$ is the pressure induced precession frequency and $\omega_{\mathcal{Q}}$ the quadrupole precession frequency. At around $R \sim 8 a_{\rm b}$, the eccentricity flattens and remains near a constant value $e_{\rm d} \sim 10^{-3}$ throughout the remainder of the disk. The flattening radius is coincident with the radius at which the disk loses coherence (Figure \ref{fig:disk_precession_circular}, second panel from the top).

As already noted from our eccentricity space-time diagrams, we observe higher disk eccentricity around the eccentric binary compared with the circular case. In the eccentric binary case, the eccentricity over nearly the entire extent of the disk follows a power law of the form $R^{-1.9}$. Our findings are in disagreement with test particle eccentricity profiles explored in \cite{MoriwakiNakagawa2004}, where the forced eccentricity follows a power law $ \propto R^{-1}$. The disk eccentricity remains close to our $R^{-1.9}$ fit throughout the outer part of the disk, before flattening near $R\sim20a_{\rm b}$.  The flattening may indicate another transition, could be caused by low resolution in the outer disk, or may have weak dependence upon the outer boundary condition.

Prograde, coherent precession of CBDs has been studied in numerical simulations (e.g. \citealt{MunozLithwick2020} for circular binaries and \citealt{Miranda2016} with equal mass, eccentric binaries), but until now has not been studied in the case of binaries with both varying mass ratios and eccentricities. In the second and fourth panel of Figure \ref{fig:disk_precession_eccentric} we plot a spacetime diagram showing the longitude of pericenter of the CBD with respect to the eccentric binary pericenter. We find that the disk around the equal mass ratio eccentric binary undergoes forced precession. However, surprisingly the CBD around the $e_{\rm b} = 0.6, \, q_{\rm b} = 0.1$ binary does not precess. Instead, the CBD is coherently locked at an angle $\lesssim -\pi/4$ with respect to the binary longitude of periapsis. Throughout our suite of simulations, we continue to find this unexpected behaviour at either low binary eccentricity and/or small mass ratios. 

\begin{figure*} 
	\centering
	\includegraphics[width=1\textwidth]{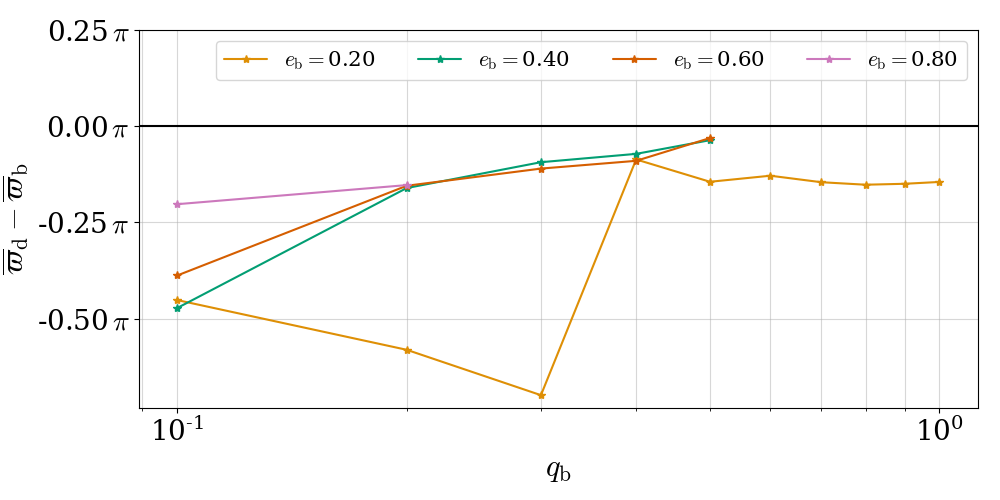}
	\caption{Angles of CBDs which lock around binaries of orbital parameters shown in red in table \ref{tab:cbdmodes}. We find that the locking angle is always negative (i.e. the disk pericenter lags `behind' the binary pericenter), and slightly increases with increasing mass ratio. We do not find any disks that are perfectly apsidally aligned with the binary.}
	\label{fig:disk_locking_angles}
\end{figure*}

In figure \ref{fig:disk_locking_angles} we show a summary of the CBD locking angles as a function of binary mass ratio. We find that when $e_{\rm b} = 0.2$, the locking angle first decreases, reaching a minimum at $q_{\rm b} = 0.3$, and then increases to a near constant value of $\overline{\varpi}_{\rm d} - \overline{\varpi}_{\rm b}  \sim -0.2 \pi$ for the remainder of the mass ratios tested. The CBD never completely apsidally aligns with the binary. In binaries with eccentricities $e_{\rm b} > 0.2$, the trend is more monotonic: $\overline{\varpi}_{\rm d} - \overline{\varpi}_{\rm b}$ consistently increases with $q_{\rm b}$, and is closest to apsidal alignment with the binary when $q_{\rm b} = 0.5$.

To give the reader some intuition for the geometry of our locked CBD-binary systems, we visualize the disk eccentricity $e_{\rm d}$ and the disk longitude of pericenter $\varpi_{\rm d}$ as a function of radius in Figure \ref{fig:fitted_ellipses_cbds}. We draw ellipses representing $[e_{\rm d},\varpi_{\rm d}]$ throughout the disk for CBDs surrounding binaries with $e_{\rm b} = 0.2$ and mass ratios ranging from $q_{\rm b} = [0.1, 0.2, 0.3, 0.5, 0.7, 1.0]$ (left to right). From top to bottom, we show the same binary-disk system $200 P_{\rm b}$ and $400 P_{\rm b}$ later. We find again that the pericenter location $\varpi_{\rm d}$ (shown as black stars) is approximately constant in the disk, and represents the bulk disk orientation by the mass-weighted, integrated global angle $\overline{\varpi}_{\rm d}$. In all cases, $\overline{\varpi}_{\rm d}$ is approximately constant for hundreds or orbits. We find that the CBD locks around binaries with eccentricity $e_{\rm b} = 0.2$ over all mass ratios tested, and find that the locking angle with respect to the binary longitude of periapsis is non-zero in all cases, meaning there is no apsidal alignment of binary and CBD.

\begin{figure*} 
	\centering
	\includegraphics[width=1\textwidth]{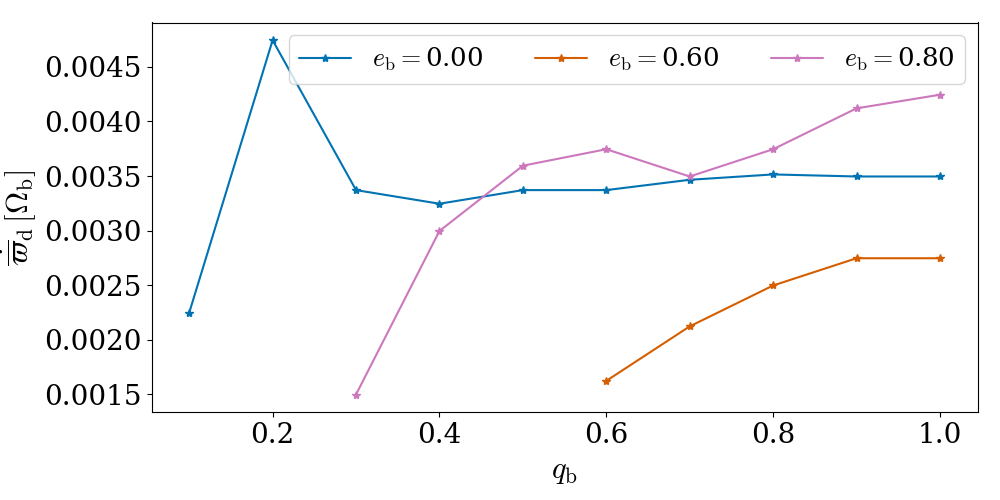}
	\caption{Precession rates of CBDs around binaries with $e_{\rm b} = 0.0$ (blue), $e_{\rm b} = 0.6$ (orange) and $e_{\rm b} = 0.8$, obtained via Fourier analysis of CBD pericenter motion. CBDs around circular binaries generally tend to precess faster at lower mass ratios, however, at higher mass ratios, the $e_{\rm b} = 0.8$ binaries induce the fastest CBD precession rates.}
	\label{fig:cbd_precession_rates}
\end{figure*}

\begin{figure} 
	\centering
	\includegraphics[width=1\columnwidth]{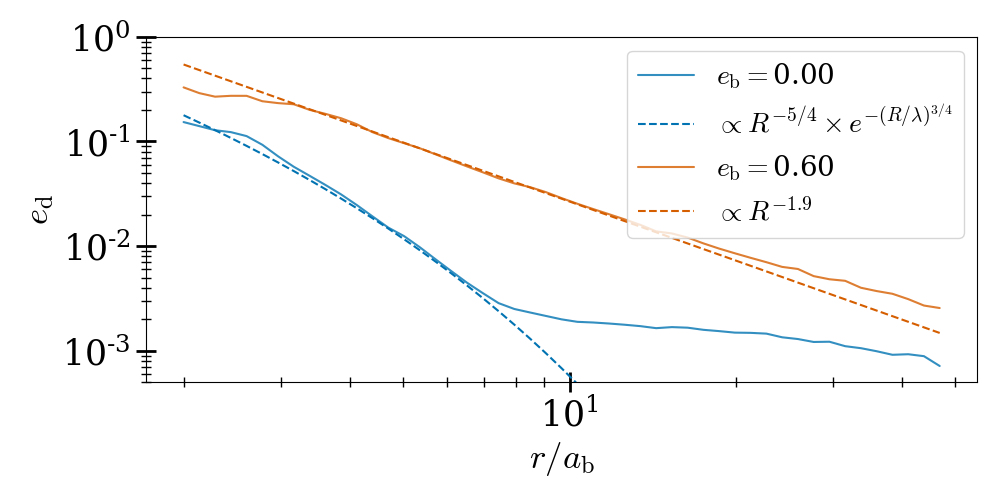}
	\caption{CBD eccentricity as a function of radius, averaged over 1000 orbits for a circular binary (blue line) and an eccentric binary with $e_{\rm b} = 0.6$ (orange line), both with mass ratio $q_{\rm b} = 0.1$. }
	\label{fig:e_vs_r}
\end{figure}

\begin{figure*} 
	\centering
	\includegraphics[width=1\textwidth]{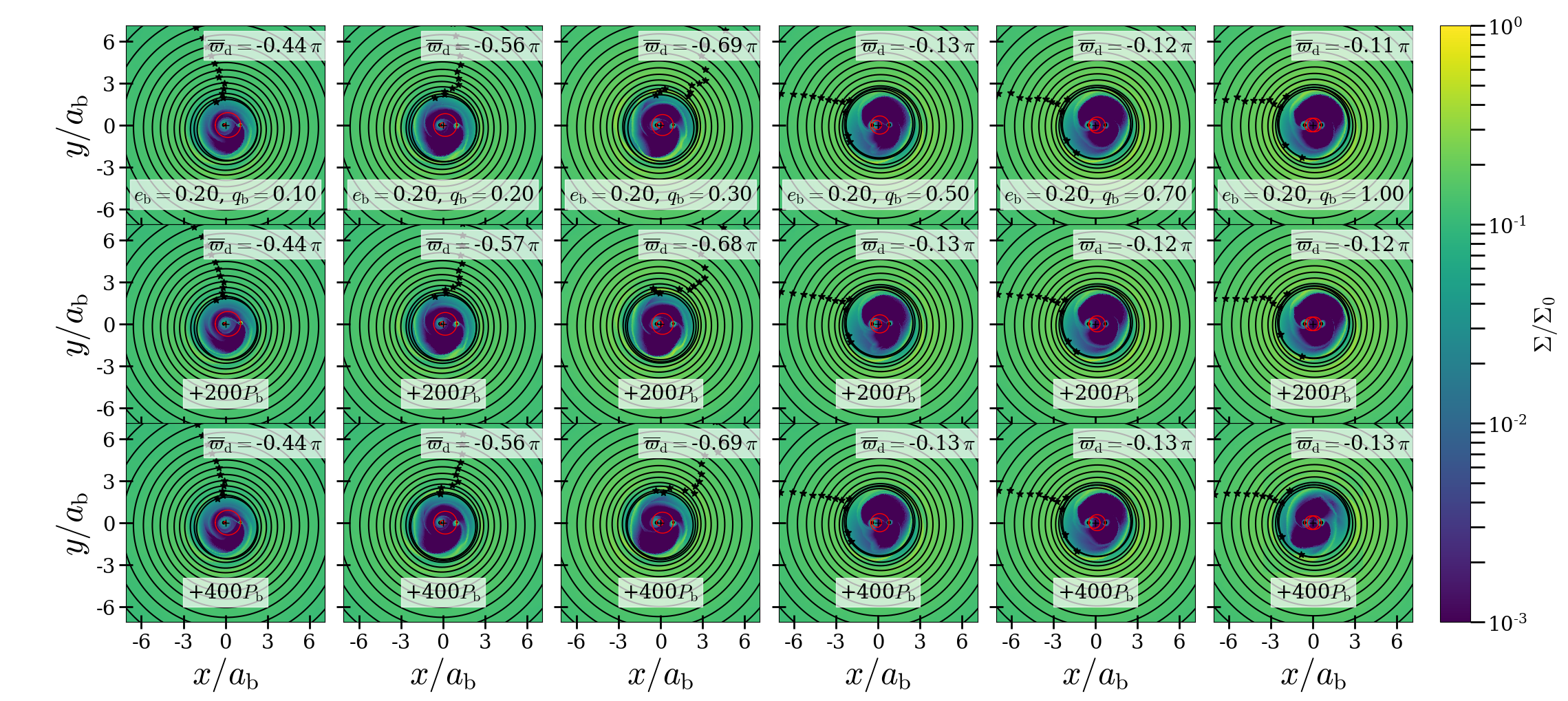}
	\caption{Visualization of locked CBDs around eccentric binaries with $e_{\rm b} = 0.2$. We draw ellipses representing the eccentricity in the disk and longitude of pericenter (black stars) at various radial bins within our CBDs. From the left most column to the right, we show mass ratios $q_{\rm b} = [0.1, 0.2, 0.3, 0.5, 0.7, 1.0]$. }
	\label{fig:fitted_ellipses_cbds}
\end{figure*}

We next compare the eccentricity and precession of disks around circular and eccentric equal mass ratio binaries, seen in the bottom two panels of Figures \ref{fig:disk_precession_circular} and \ref{fig:disk_precession_eccentric}. In both cases, the CBD eccentricity is substantial ($e_{\rm b} \sim 0.3$) and fluctuates on timescales of the CBD precession frequency.
We find the precession frequency around the $e_{\rm b} = 0.6$, equal mass binary to be lower than that of the circular binary (though note in Figure \ref{fig:cbd_precession_rates} that this is not always the case).  If CBD precession was fully explained by the varying quadrupole potential, equation (\ref{eqn:quadrupole_frequency}) would predict the disk to precess faster around an eccentric binary of the same mass ratio. 
In figure \ref{fig:cbd_precession_rates} we show the precession rates of all CBDs in our simulations, around circular binaries (in blue) and around binaries with eccentricities $e_{\rm b} = 0.6$ (orange) and $e_{\rm b} = 0.8$ (pink) as a function of mass ratio. We omit the $e_{\rm b} = 0.4$ case, since the CBD precession remains irregular over the 10,000 orbits tested. A Fourier analysis of the disk pericenter precession does not return a clear signal, and as such the precession rate cannot be characterized by a single value. However, Figure \ref{fig:cbd_precession_rates} shows that CBD precession rates increase as binary eccentricity grows, exceeding the precession rate around a circular orbit when $q_{\rm b} > 0.4$.  Our findings are generally consistent with Figure 3 in \cite{DorazioDuffell2021}, in that precession around circular, $q_{\rm b} = 1.0$ binaries is faster than around binaries with $e_{\rm b} \sim 0.6$, and that disk precession speeds up slightly as binary eccentricity grows beyond $e_{\rm b} \gtrsim 0.6$. We suggest that further theoretical follow-up study is needed to explain the non-monotonic relationship between CBD precession rates and binary eccentricity.

\setlength{\arrayrulewidth}{0.4mm}
\setlength{\tabcolsep}{10pt}
\arrayrulecolor[HTML]{480607}
\begin{table}
	\centering
	\begin{tabular}{ |c|c|c|c|c|c|}
		\hline
		 \diagbox{$q_{\rm b}$}{$e_{\rm b}$} & $0.0$ & $0.2$ & $0.4$  &  $0.6$ & $0.8$ \\
		\hline
		 0.1 &\cellcolor[HTML]{FFD700}  $\mathcal{P}$ &  \cellcolor[HTML]{AA0044} $\mathcal{L}$ & \cellcolor[HTML]{AA0044} $\mathcal{L}$ & \cellcolor[HTML]{AA0044} $\mathcal{L}$  & \cellcolor[HTML]{AA0044} $\mathcal{L}$ \\
		\hline
		0.2 &\cellcolor[HTML]{FFD700}  $\mathcal{P}$ & \cellcolor[HTML]{AA0044} $\mathcal{L}$  & \cellcolor[HTML]{AA0044} $\mathcal{L}$ & \cellcolor[HTML]{AA0044} $\mathcal{L}$ & \cellcolor[HTML]{AA0044} $\mathcal{L}$ \\
		\hline
		 0.3 & \cellcolor[HTML]{FFD700} $\mathcal{P}$ & \cellcolor[HTML]{AA0044} $\mathcal{L}$ & \cellcolor[HTML]{AA0044} $\mathcal{L}$ &\cellcolor[HTML]{AA0044} $\mathcal{L}$ &  \cellcolor[HTML]{007FFF} $\mathcal{F}$ \\
		\hline
		 0.4 & \cellcolor[HTML]{FFD700} $\mathcal{P}$ & \cellcolor[HTML]{AA0044} $\mathcal{L}$ & \cellcolor[HTML]{AA0044} $\mathcal{L}$ &\cellcolor[HTML]{AA0044} $\mathcal{L}$ & \cellcolor[HTML]{007FFF} $\mathcal{F}$\\
		\hline
		0.5 &  \cellcolor[HTML]{FFD700} $\mathcal{P}$ & \cellcolor[HTML]{AA0044} $\mathcal{L}$  & \cellcolor[HTML]{AA0044} $\mathcal{L}$ &\cellcolor[HTML]{AA0044} $\mathcal{L}$& \cellcolor[HTML]{007FFF} $\mathcal{F}$\\
		\hline
		0.6 & \cellcolor[HTML]{FFD700}  $\mathcal{P}$ & \cellcolor[HTML]{AA0044} $\mathcal{L}$  & \cellcolor[HTML]{007FFF} $\mathcal{F}$ & \cellcolor[HTML]{007FFF} $\mathcal{F}$ & \cellcolor[HTML]{007FFF} $\mathcal{F}$ \\
	    \hline
		0.7 & \cellcolor[HTML]{FFD700}  $\mathcal{P}$ & \cellcolor[HTML]{AA0044} $\mathcal{L}$ & \cellcolor[HTML]{007FFF} $\mathcal{F}$ & \cellcolor[HTML]{007FFF} $\mathcal{F}$ & \cellcolor[HTML]{007FFF} $\mathcal{F}$ \\
		\hline
		0.8 & \cellcolor[HTML]{FFD700}  $\mathcal{P}$ & \cellcolor[HTML]{AA0044} $\mathcal{L}$ & \cellcolor[HTML]{007FFF} $\mathcal{F}$ & \cellcolor[HTML]{007FFF} $\mathcal{F}$ & \cellcolor[HTML]{007FFF} $\mathcal{F}$ \\
		\hline
		0.9 & \cellcolor[HTML]{FFD700} $\mathcal{P}$ & \cellcolor[HTML]{AA0044} $\mathcal{L}$ & \cellcolor[HTML]{007FFF} $\mathcal{F}$ & \cellcolor[HTML]{007FFF} $\mathcal{F}$ & \cellcolor[HTML]{007FFF} $\mathcal{F} $ \\
		\hline
		1.0 & \cellcolor[HTML]{FFD700} $\mathcal{P}$ & \cellcolor[HTML]{AA0044} $\mathcal{L}$  & \cellcolor[HTML]{007FFF} $\mathcal{F}$  & \cellcolor[HTML]{007FFF} $\mathcal{F}$ & \cellcolor[HTML]{007FFF} $\mathcal{F}$  \\
		\hline
	\end{tabular}
	\caption{CBD modes as a function of $q_{\rm b}$ and $e_{\rm b}$. In this table, $\mathcal{P}$ indicates `freely precessing', $\mathcal{L}$ `locked', and $\mathcal{F}$  represents CBDs undergoing `forced precession'.}
  \label{tab:cbdmodes}
\end{table}
Throughout our suite of simulations we find that the response of the CBD to the binary falls into one of three categories: CBDs can either precess freely around circular binaries, remain locked at an angle with respect to the binary periapsis, or undergo forced precession around the binary. Whether a disk precesses or locks depends on the parameters of the central binary. Locked disks, as shown in Figures \ref{fig:disk_precession_eccentric} and \ref{fig:fitted_ellipses_cbds}, are found at either low eccentricity or low mass ratio, while CBDs tend to undergo forced precession around binaries of either high mass ratio or high eccentricity. We summarize our results in Table  \ref{tab:cbdmodes} as a function of $q_{\rm b}$ and $e_{\rm b}$: Free precession (yellow $\mathcal{P}$) is seen exclusively in circular binaries. Locked disks (red $\mathcal{L}$) occur at the lowest eccentricity tested, or at low mass ratios in the remaining eccentricities. Forced precession (blue $\mathcal{F}$) is observed in CBDs around binaries with moderate-high mass ratios and high eccentricity. The locking and precession regimes we find for $q_{\rm b} = 1.0$ binaries are similar to results in \cite{DorazioDuffell2021}, while some differences in our findings are discussed in section \ref{sec:discussion}.

\subsection{Accretion Variability}
\begin{figure*} 
	\centering
	\includegraphics[width=1.0\textwidth]{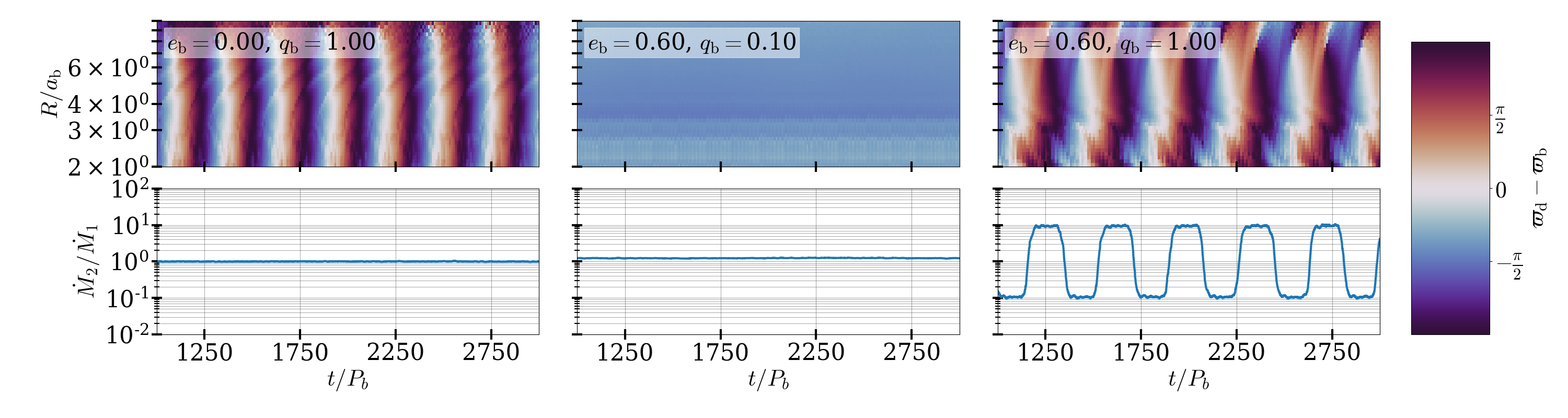}
	\caption{Three CBD response regimes: free precession, locked CBD  and forced precession (top panels), and their respective effect on preferential accretion rates (bottom panels). Left column: Circular binary with equal mass ratio. The disk around the circular binary freely precesses without causing any effect on the accretion behaviour. Middle column:  Eccentric low mass ratio binary, with $e_{\rm b} = 0.6, \, q_{\rm b} = 0.1$, and a locked CBD. The constant value of the angle $\varpi = \varpi_{\rm d} - \varpi_{\rm b}$ confirms that no precession takes place over the period of $2000 \, P_{\rm b}$ shown. The preferential accretion rate is shown in the bottom panel, and remains constant over the same period. Right column: Eccentric equal mass binary, with $e_{\rm b} = 0.6, \, q_{\rm b} = 1.0$, and forced precession of the CBD. The forced precession is associated with 2 orders of magnitude fluctuations in preferential accretion rates.}
	\label{fig:preferential_accretion_precession}
\end{figure*}

Similar to previous work on eccentric binaries \citep{Dunhill2015, Munoz2016, DorazioDuffell2021}, we find that the forced precession of CBDs around eccentric binaries is associated with accretion variability. In Figure \ref{fig:preferential_accretion_precession} we contrast the accretion behaviour of our three disk regimes: free precession around a circular, equal mass binary (left), disk locking around a binary with parameters $e_{\rm b} = 0.6$,  $q_{\rm b} = 0.1$ (middle), and forced precession around  $e_{\rm b} = 0.6$,  $q_{\rm b} = 1.0$ (right).

 We find that the preferential accretion rate is constant over thousands of orbits when, \begin{enumerate*}[label=(\roman*)] \item the binary orbit is circular (see panels on the left), \textit{or} \item the disk is locked, as is the case for the eccentric, low mass ratio binary in the middle panel
 \end{enumerate*}. However, in the case of forced disk precessing around eccentric binaries (right), we find that the preferential accretion rate fluctuates by an order of magnitude around its mean. The period of the fluctuation exactly coincides with that of the forced precession, and $\lambda$ appears to peak at maximum misalignment (when $\varpi \sim \pi$), when the pericenter of the disk is closest to the secondary. Whether $\lambda$ peaks at $\varpi \sim \pi$ or $\varpi \sim 0$ just depends on the labeling of ``primary" vs ``secondary", which is an arbitrary choice in the case of an equal mass binary.

\begin{figure*}
	\centering
	\includegraphics[width=1.0\textwidth]{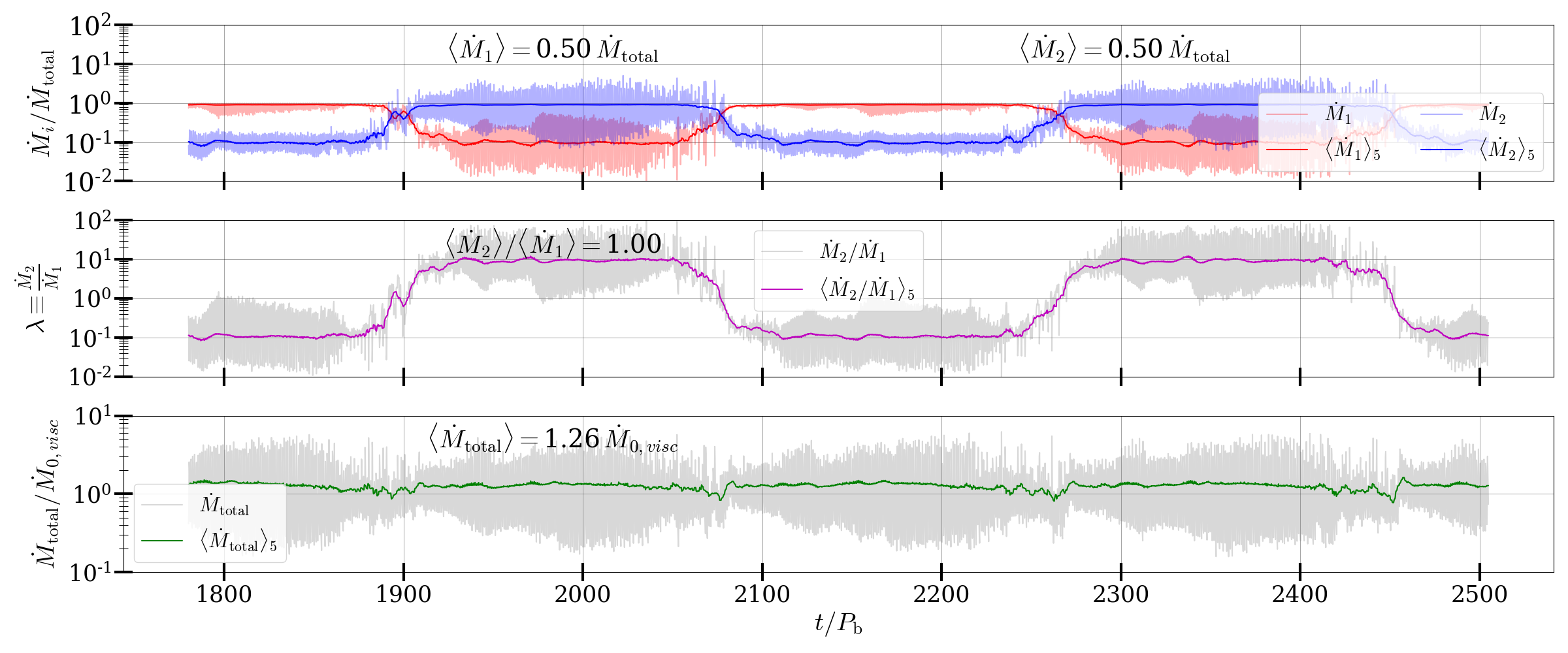}
	\caption{Accretion variability of a $q_{\rm b} = 1.0$, $e_{\rm b} = 0.6$ binary over 700 $P_{\rm b}$. The top panel shows the individual accretion rates of primary (red) and secondary (blue). The middle panel is the preferential accretion rate $\lambda$. The bottom panel shows the total accretion rate. All three panels exhibit accretion variability on the CBD precession timescale at a period $\sim 200 \, P_{\rm b}$. The preferential accretion rate is exactly 1 when averaged over a few precession timescales, as required by symmetry arguments.}
	\label{fig:individual_pref_total_accretion}
\end{figure*}

In the case of all equal mass binaries, symmetry arguments require that the preferential accretion rate average out to exactly 1. For eccentric binaries, the azimuthal symmetry of the binary orbit is broken, resulting in temporarily boosted accretion rates onto the binary component orbiting closer to the precessing pericenter of the disk. Although the CBD precesses around all of our circular binaries, the symmetry of the circular binary orbit prevents any variability, as can be seen in the constant accretion rates on the left hand panels in Figure \ref{fig:preferential_accretion_precession}. In eccentric binaries  however (right hand panels), the symmetry breaking results in periodically fluctuating preferential accretion rates (bottom right panel). The periodicity of the accretion variability occurs on timescales of the forced CBD precession. Symmetry arguments still require that the preferential accretion rates onto eccentric, equal mass binaries converge to exactly 1, when averaged over timescales longer than the disk precession period.

We show that this is  indeed the case in Figure \ref{fig:individual_pref_total_accretion}, for the $e_{\rm b}= 0.6$, $q_{\rm b} = 1.0$ case. The top panel shows the individual accretion rates of the ``primary" (red) and ``secondary" (blue), switching accretion preference every  $\sim 200 P_{\rm b}$. The middle panel shows the value of $\lambda$ over time, indicating the same pattern as in the bottom right panel of Figure \ref{fig:preferential_accretion_precession}. As required, we find that $\lambda = 1$ (exactly) when averaged out over integer multiples of the disk precession timescale. The bottom panel of Figure \ref{fig:individual_pref_total_accretion} shows the total accretion rate onto the binary. The raw values (grey line) indicate strong accretion fluctuation, which is also shown averaged out over 5 orbits (green line), where periodicity is still discernible.

We also find long term accretion variability at other mass ratios and eccentricities, whenever there is forced precession of the CBD around the binary. The binary parameters for which this occurs can be found in  Table \ref{tab:cbdmodes}.

\section{Summary and Discussion}
\label{sec:discussion}
We have presented a suite of high resolution 2d hydrodynamic simulations, which measure the binary preferential accretion rates as a function of both mass ratio and eccentricity. We have shown that preferential accretion varies not only as a function of mass ratio, as was previously known, but also non-monotonically as a function of eccentricity (Figure \ref{fig:eccentric_pref_acc}). At low eccentricity, the preferential accretion onto the secondary is enhanced when compared with circular binaries, and at high eccentricity ($e_{\rm b} > 0.6$), preferential accretion is damped. Using our models we have calculated the timescales over which mass ratios of binaries equalize due to preferential accretion: accreting at 10\% of the Eddington limit, a binary with initial mass ratio $q_{\rm b} = 0.1$ will take between $500$~Myr (low eccentricity) to $\sim 3$~Gyr (high eccentricity), as shown in Figure \ref{fig:timescale_to_equalizing}. These numbers are scale-free and would apply to both stellar mass binaries and MBHBs, however, our choice of disk aspect ratio ($h = 0.1$), viscosity model and isothermal equation of state must be taken into account when interpreting our results. 
The near order of magnitude difference in mass ratio equalization timescales suggests that in binaries where accretion disks are present for significant timescales, mass ratio distributions may shift systematically as a function of binary eccentricity. 

However, the amplitude of this effect may be limited by circumbinary disk lifetimes: AGN disks are expected to persist for around $\sim 10^8\rm{yrs}$ (e.g. \citealt{Yu2002}), and up to $10^9 \, \rm{yrs}$ given certain models \citep{Marconi2004}. This timescale may be too short to grow binaries from $q_{\rm b} = 0.1 \rightarrow 1$, except for binaries with low-moderate eccentricities $e_{\rm b} = 0.2 - 0.4$. Higher eccentricity models ($e_{\rm b} > 0.4$) may require accretion rates exceeding $10\%$ Eddington, or multiple AGN disk lifetimes to produce a significant increase in MBHB mass ratio distributions. Stellar disks on the other hand last for only $\sim 10^7  \, \rm{yrs}$ \citep{Li2016}, and are therefore less likely to facilitate significant mass ratio growth, unless stellar binaries accrete at or above the Eddington limit for extended periods of time.

Nevertheless, correlations between mass ratio and eccentricity have been observed in binary star systems by \cite{Halbwachs2002}, who found that stellar binaries with near equal mass ratios also tend to have lower eccentricities, while more eccentric binaries have lower mass ratios.
Similarly, \cite{Moe2017} found that low eccentricities were associated with an excess of ``twins", or high mass ratios, in their sample of stellar binaries with small separations ($a_b \lesssim 0.4$ AU).
The above observed population statistic mirrors the preferential accretion behaviour of binaries we have found in this work, though more studies with larger sample sizes are needed to robustly draw a connection with CBD accretion.

We note that the timescales presented in Figure \ref{fig:timescale_to_equalizing} assume that binaries keep a constant eccentricity as their mass ratios grow from their initial value $q_{b,0}$ towards 1. However, recent simulations of binaries with equal mass ratios have suggested that interaction with the CBD drives the binary eccentricity to a critical, steady state value near $e_{\rm crit} \sim 0.4$ \citep{Cuadra2009, Munoz2020, Zrake2021, DorazioDuffell2021}. For smaller mass ratios, \cite{Roedig2011} found that at $q_{\rm b} = 1/3$, the critical eccentricity of the binary is higher, between $0.6-0.8$. This hints at the possibility that the steady state eccentricity of binaries in CBDs is a function of mass ratio, and that both eccentricity and mass ratio could evolve as the binary accretes.
If the eccentricity evolution is faster than the change in mass ratio, the  \textit{mass ratio growth timescales could vary over time as mass ratios and eccentricities change} as a result of CBD accretion and eccentricity saturation. If the timescales of eccentricity saturation and the relationship between steady state eccentricity and mass ratio are investigated further, mass ratio growth timescales for binaries can be obtained even more accurately by interpolating our preferential accretion models as shown in Figure  \ref{fig:interp_lambda}, 
 though further simulations would be helpful to better understand the effect of disk scale height and viscosity treatment on these results.

We have found that the response of the CBD to the inner binary can fall into one of three regimes: free precession, disk periapsis locking or forced precession. Which behaviour is observed depends on the binary parameters, and can be found in Table \ref{tab:cbdmodes}. We have demonstrated that the locking angle between CBD and binary periapsis is non-zero and depends on both mass ratio and eccentricity (see figure \ref{fig:disk_locking_angles}). We find no instance of perfect apsidal alignment between binary and disk, however the locking angle between disk and binary tends towards apsidal alignment as mass ratios increase. 

Our result of non-apsidally aligned locked CBDs adds a new form of binary-disk interaction to previous findings in \cite{Miranda2016}, \cite{Thun2017} and \cite{DorazioDuffell2021}, who find either apsidal alignment or forced precession of disks around binaries. 
\cite{DorazioDuffell2021} presented a densely sampled parameter study in equal-mass binaries with eccentricities ranging continuously from $0.0 \rightarrow 0.9$, and found that the disk undergoes a transition from apsidally locked between $0.2 \lesssim e_{\rm b} \lesssim 0.4$ to precessing at larger eccentricities. 
\cite{Thun2017} carried out a parameter study over eccentricity ranging from $e_{\rm b}: [0.0 \rightarrow 0.64]$ with fixed mass ratio $q_{\rm b} = 0.29$, and a parameter study over mass ratio ranging from $q_{\rm b}: [0.1 \rightarrow 1.0]$ with fixed eccentricity $e_{\rm b} = 0.16$, and similarly to our work compared the precession of the CBD as a function of those varying parameters. They observed no `stand-still' disks (which we call `locked' here) as long as the inner boundary of their computational domain was chosen small enough, and attributed similarly locked disks found by \cite{Miranda2016} to the inner edge of their computational domain. However, in our simulation setup with \texttt{Arepo} we do not require an inner boundary, and still find non-precessing CBDs in parts of our parameter study, particularly around low mass-ratio binaries and at very low eccentricity. \cite{Miranda2016} and \cite{DorazioDuffell2021} find a locked disk at equal mass ratio and $e_{\rm b} = 0.2$, in agreement with our study, however their disks were apsidally aligned with the binary. \cite{Miranda2016} and \cite{DorazioDuffell2021} also find a locked disk at $q_{\rm b} = 1, \, e_{\rm b}=0.4$, where our simulations instead find forced precession, albeit at irregular precession rates. 
The disagreement might be due to differences in the disk setup: \cite{Miranda2016} and  \cite{DorazioDuffell2021} evolve their binaries in an infinite disk, with continuous mass inflow at the outer boundary, whereas our disk is finite and viscously spreading, but further study is needed to confirm the cause of the discrepancy.

Our results give tantalizing prospects for upcoming transient surveys of variable AGN (e.g. \citealt{Charisi2021}) and population studies of binaries in general, by predicting accretion variability signatures, and effects of long term binary accretion on mass ratio distributions. While our results are scale-free and apply to stellar binaries as well as supermassive black holes, we point out that Mach numbers are expected to vary depending on the scale of the binary. AGN disks are typically assumed to be much thinner than our fiducial scale height $h = 0.1$, and the accretion behaviour and CBD alignment in such systems may differ as a result. Further studies could generalize our results by including varying disk scale heights (similar to e.g. \cite{Tiede2020}, but with non equal mass ratios), and angular momentum transport due to magnetorotational instability-driven turbulence \citep{BalbusHawley1991}. 

In the case of MBHBs, CBDs are replenished by inflowing gas funnelled towards the galactic center at random angles with respect to the binary plane, resulting in inclined CBDs. Future 3d simulations should include the effect of inclination angles on preferential accretion rates.

\section{Acknowledgements}
\label{sec:ack}
We are grateful to Zachary Murray and Paul Duffell for helpful discussions.
This research made use of \texttt{SciPy} \citep{2020SciPy-NMeth} and \texttt{NumPy} \citep{vanderWalt2011}.
\texttt{Seaborn} \citep{seaborn} and \texttt{MATPLOTLIB} \citep{Hunter2007} were used to generate figures.
RW is supported by the Natural Sciences and Engineering Research Council of Canada (NSERC), funding reference \#CITA 490888-16.

\section{Data Availability}
The data underlying this article will be shared on reasonable request to the corresponding author.


\twocolumn

\bibliographystyle{mnras}
\bibliography{mybib} 

\begin{thebibliography}{}
\makeatletter
\relax
\def\mn@urlcharsother{\let\do\@makeother \do\$\do\&\do\#\do\^\do\_\do\%\do\~}
\def\mn@doi{\begingroup\mn@urlcharsother \@ifnextchar [ {\mn@doi@}
  {\mn@doi@[]}}
\def\mn@doi@[#1]#2{\def\@tempa{#1}\ifx\@tempa\@empty \href
  {http://dx.doi.org/#2} {doi:#2}\else \href {http://dx.doi.org/#2} {#1}\fi
  \endgroup}
\def\mn@eprint#1#2{\mn@eprint@#1:#2::\@nil}
\def\mn@eprint@arXiv#1{\href {http://arxiv.org/abs/#1} {{\tt arXiv:#1}}}
\def\mn@eprint@dblp#1{\href {http://dblp.uni-trier.de/rec/bibtex/#1.xml}
  {dblp:#1}}
\def\mn@eprint@#1:#2:#3:#4\@nil{\def\@tempa {#1}\def\@tempb {#2}\def\@tempc
  {#3}\ifx \@tempc \@empty \let \@tempc \@tempb \let \@tempb \@tempa \fi \ifx
  \@tempb \@empty \def\@tempb {arXiv}\fi \@ifundefined
  {mn@eprint@\@tempb}{\@tempb:\@tempc}{\expandafter \expandafter \csname
  mn@eprint@\@tempb\endcsname \expandafter{\@tempc}}}

\bibitem[\protect\citeauthoryear{Andrews, Wilner, Hughes, Qi  \&
  Dullemond}{Andrews et~al.}{2009}]{Andrews2009}
Andrews S.~M.,  Wilner D.~J.,  Hughes A.~M.,  Qi C.,   Dullemond C.~P.,  2009,
  \mn@doi [The Astrophysical Journal] {10.1088/0004-637x/700/2/1502}, 700, 1502

\bibitem[\protect\citeauthoryear{{Artymowicz}}{{Artymowicz}}{1983}]{Artymowicz1983}
{Artymowicz} P.,  1983, Postepy Astronomii Krakow, \href
  {https://ui.adsabs.harvard.edu/abs/1983PoAst..31...19A} {31, 19}

\bibitem[\protect\citeauthoryear{{Artymowicz} \& {Lubow}}{{Artymowicz} \&
  {Lubow}}{1996}]{ArtymowiczLubow1996}
{Artymowicz} P.,  {Lubow} S.~H.,  1996, \mn@doi [\apjl] {10.1086/310200}, \href
  {https://ui.adsabs.harvard.edu/abs/1996ApJ...467L..77A} {467, L77}

\bibitem[\protect\citeauthoryear{{Balbus} \& {Hawley}}{{Balbus} \&
  {Hawley}}{1991}]{BalbusHawley1991}
{Balbus} S.~A.,  {Hawley} J.~F.,  1991, \mn@doi [\apj] {10.1086/170270}, \href
  {https://ui.adsabs.harvard.edu/abs/1991ApJ...376..214B} {376, 214}

\bibitem[\protect\citeauthoryear{Bate, Bonnell  \& Bromm}{Bate
  et~al.}{2002}]{Bate2002}
Bate M.~R.,  Bonnell I.~A.,   Bromm V.,  2002, \mn@doi [Monthly Notices of the
  Royal Astronomical Society] {10.1046/j.1365-8711.2002.05775.x}, 336, 705

\bibitem[\protect\citeauthoryear{Bortolas, Franchini, Bonetti  \&
  Sesana}{Bortolas et~al.}{2021}]{Bortolas2021}
Bortolas E.,  Franchini A.,  Bonetti M.,   Sesana A.,  2021, \mn@doi [The
  Astrophysical Journal Letters] {10.3847/2041-8213/ac1c0c}, 918, L15

\bibitem[\protect\citeauthoryear{Charisi, Taylor, Runnoe, Bogdanovic  \&
  Trump}{Charisi et~al.}{2021}]{Charisi2021}
Charisi M.,  Taylor S.~R.,  Runnoe J.,  Bogdanovic T.,   Trump J.~R.,  2021, ]
  {10.48550/ARXIV.2110.14661}

\bibitem[\protect\citeauthoryear{Cuadra, Armitage, Alexander  \&
  Begelman}{Cuadra et~al.}{2009}]{Cuadra2009}
Cuadra J.,  Armitage P.~J.,  Alexander R.~D.,   Begelman M.~C.,  2009, \mn@doi
  [Monthly Notices of the Royal Astronomical Society]
  {10.1111/j.1365-2966.2008.14147.x}, 393, 1423

\bibitem[\protect\citeauthoryear{D'Orazio \& Duffell}{D'Orazio \&
  Duffell}{2021}]{DorazioDuffell2021}
D'Orazio D.~J.,  Duffell P.~C.,  2021, \mn@doi [The Astrophysical Journal]
  {10.3847/2041-8213/ac0621}, 914, L21

\bibitem[\protect\citeauthoryear{{D'Orazio}, {Haiman}  \&
  {MacFadyen}}{{D'Orazio} et~al.}{2013}]{DOrazio2013}
{D'Orazio} D.~J.,  {Haiman} Z.,   {MacFadyen} A.,  2013, \mn@doi [\mnras]
  {10.1093/mnras/stt1787}, \href
  {https://ui.adsabs.harvard.edu/abs/2013MNRAS.436.2997D} {436, 2997}

\bibitem[\protect\citeauthoryear{{D'Orazio}, {Haiman}, {Duffell}, {MacFadyen}
  \& {Farris}}{{D'Orazio} et~al.}{2016}]{DOrazio2016}
{D'Orazio} D.~J.,  {Haiman} Z.,  {Duffell} P.,  {MacFadyen} A.,   {Farris} B.,
  2016, \mn@doi [\mnras] {10.1093/mnras/stw792}, \href
  {https://ui.adsabs.harvard.edu/abs/2016MNRAS.459.2379D} {459, 2379}

\bibitem[\protect\citeauthoryear{Dittmann \& Ryan}{Dittmann \&
  Ryan}{2021}]{Dittmann2021}
Dittmann A.,  Ryan G.,  2021, Preventing Anomalous Torques in Circumbinary
  Accretion Simulations (\mn@eprint {} {arXiv:2102.05684})

\bibitem[\protect\citeauthoryear{Duffell, D'Orazio, Derdzinski, Haiman,
  MacFadyen, Rosen  \& Zrake}{Duffell et~al.}{2020}]{Duffell2020}
Duffell P.~C.,  D'Orazio D.,  Derdzinski A.,  Haiman Z.,  MacFadyen A.,  Rosen
  A.~L.,   Zrake J.,  2020, \mn@doi [The Astrophysical Journal]
  {10.3847/1538-4357/abab95}, 901, 25

\bibitem[\protect\citeauthoryear{Dunhill, Cuadra  \& Dougados}{Dunhill
  et~al.}{2015}]{Dunhill2015}
Dunhill A.~C.,  Cuadra J.,   Dougados C.,  2015, \mn@doi [Monthly Notices of
  the Royal Astronomical Society] {10.1093/mnras/stv284}, 448, 3545

\bibitem[\protect\citeauthoryear{Farris, Duffell, MacFadyen  \& Haiman}{Farris
  et~al.}{2014}]{Farris2014}
Farris B.~D.,  Duffell P.,  MacFadyen A.~I.,   Haiman Z.,  2014, \mn@doi [The
  Astrophysical Journal] {10.1088/0004-637x/783/2/134}, 783, 134

\bibitem[\protect\citeauthoryear{{Gerosa}, {Veronesi}, {Lodato}  \&
  {Rosotti}}{{Gerosa} et~al.}{2015}]{Gerosa2015}
{Gerosa} D.,  {Veronesi} B.,  {Lodato} G.,   {Rosotti} G.,  2015, \mn@doi
  [\mnras] {10.1093/mnras/stv1214}, \href
  {https://ui.adsabs.harvard.edu/abs/2015MNRAS.451.3941G} {451, 3941}

\bibitem[\protect\citeauthoryear{Halbwachs, Mayor, Udry  \& Arenou}{Halbwachs
  et~al.}{2002}]{Halbwachs2002}
Halbwachs J.~L.,  Mayor M.,  Udry S.,   Arenou F.,  2002, \mn@doi [Astronomy
  {\&} Astrophysics] {10.1051/0004-6361:20021507}, 397, 159

\bibitem[\protect\citeauthoryear{Hartmann, Calvet, Gullbring  \&
  Alessio}{Hartmann et~al.}{1998}]{Hartmann1998}
Hartmann L.,  Calvet N.,  Gullbring E.,   Alessio P.~D.,  1998, \mn@doi [The
  Astrophysical Journal] {10.1086/305277}, 495, 385

\bibitem[\protect\citeauthoryear{Hunter}{Hunter}{2007}]{Hunter2007}
Hunter J.~D.,  2007, \mn@doi [Computing in Science {\&} Engineering]
  {10.1109/mcse.2007.55}, 9, 90

\bibitem[\protect\citeauthoryear{Kelley, Haiman, Sesana  \& Hernquist}{Kelley
  et~al.}{2019}]{Kelley2019}
Kelley L.~Z.,  Haiman Z.,  Sesana A.,   Hernquist L.,  2019, \mn@doi [Monthly
  Notices of the Royal Astronomical Society] {10.1093/mnras/stz150}, 485, 1579

\bibitem[\protect\citeauthoryear{Li \& Xiao}{Li \& Xiao}{2016}]{Li2016}
Li M.,  Xiao L.,  2016, \mn@doi [The Astrophysical Journal]
  {10.3847/0004-637x/820/1/36}, 820, 36

\bibitem[\protect\citeauthoryear{MacFadyen \& Milosavljevi{\'{c}}}{MacFadyen \&
  Milosavljevi{\'{c}}}{2008}]{MacFadyen2008}
MacFadyen A.~I.,  Milosavljevi{\'{c}} M.,  2008, \mn@doi [The Astrophysical
  Journal] {10.1086/523869}, 672, 83

\bibitem[\protect\citeauthoryear{Marconi, Risaliti, Gilli, Hunt, Maiolino  \&
  Salvati}{Marconi et~al.}{2004}]{Marconi2004}
Marconi A.,  Risaliti G.,  Gilli R.,  Hunt L.~K.,  Maiolino R.,   Salvati M.,
  2004, \mn@doi [Monthly Notices of the Royal Astronomical Society]
  {10.1111/j.1365-2966.2004.07765.x}, 351, 169

\bibitem[\protect\citeauthoryear{Miranda, Mu{\~{n}}oz  \& Lai}{Miranda
  et~al.}{2016}]{Miranda2016}
Miranda R.,  Mu{\~{n}}oz D.~J.,   Lai D.,  2016, \mn@doi [Monthly Notices of
  the Royal Astronomical Society] {10.1093/mnras/stw3189}, 466, 1170

\bibitem[\protect\citeauthoryear{Moe \& Stefano}{Moe \&
  Stefano}{2017}]{Moe2017}
Moe M.,  Stefano R.~D.,  2017, \mn@doi [The Astrophysical Journal Supplement
  Series] {10.3847/1538-4365/aa6fb6}, 230, 15

\bibitem[\protect\citeauthoryear{{Moriwaki} \& {Nakagawa}}{{Moriwaki} \&
  {Nakagawa}}{2004}]{MoriwakiNakagawa2004}
{Moriwaki} K.,  {Nakagawa} Y.,  2004, \mn@doi [\apj] {10.1086/421342}, \href
  {https://ui.adsabs.harvard.edu/abs/2004ApJ...609.1065M} {609, 1065}

\bibitem[\protect\citeauthoryear{{Mu{\~n}oz} \& {Lithwick}}{{Mu{\~n}oz} \&
  {Lithwick}}{2020}]{MunozLithwick2020}
{Mu{\~n}oz} D.~J.,  {Lithwick} Y.,  2020, \mn@doi [\apj]
  {10.3847/1538-4357/abc74c}, \href
  {https://ui.adsabs.harvard.edu/abs/2020ApJ...905..106M} {905, 106}

\bibitem[\protect\citeauthoryear{{Mu{\~n}oz}, {Springel}, {Marcus},
  {Vogelsberger}  \& {Hernquist}}{{Mu{\~n}oz} et~al.}{2013}]{Munoz2013}
{Mu{\~n}oz} D.~J.,  {Springel} V.,  {Marcus} R.,  {Vogelsberger} M.,
  {Hernquist} L.,  2013, \mn@doi [\mnras] {10.1093/mnras/sts015}, \href
  {https://ui.adsabs.harvard.edu/abs/2013MNRAS.428..254M} {428, 254}

\bibitem[\protect\citeauthoryear{{Mu{\~n}oz}, {Lai}, {Kratter}  \&
  {Miranda}}{{Mu{\~n}oz} et~al.}{2020}]{Munoz2020}
{Mu{\~n}oz} D.~J.,  {Lai} D.,  {Kratter} K.,   {Miranda} R.,  2020, \mn@doi
  [\apj] {10.3847/1538-4357/ab5d33}, \href
  {https://ui.adsabs.harvard.edu/abs/2020ApJ...889..114M} {889, 114}

\bibitem[\protect\citeauthoryear{Mu{\~{n}}oz \& Lai}{Mu{\~{n}}oz \&
  Lai}{2016}]{Munoz2016}
Mu{\~{n}}oz D.~J.,  Lai D.,  2016, \mn@doi [The Astrophysical Journal]
  {10.3847/0004-637x/827/1/43}, 827, 43

\bibitem[\protect\citeauthoryear{Mu{\~{n}}oz, Miranda  \& Lai}{Mu{\~{n}}oz
  et~al.}{2019}]{Munoz2019b}
Mu{\~{n}}oz D.~J.,  Miranda R.,   Lai D.,  2019, \mn@doi [The Astrophysical
  Journal] {10.3847/1538-4357/aaf867}, 871, 84

\bibitem[\protect\citeauthoryear{{Pakmor}, {Springel}, {Bauer}, {Mocz},
  {Munoz}, {Ohlmann}, {Schaal}  \& {Zhu}}{{Pakmor} et~al.}{2016}]{Pakmor2016}
{Pakmor} R.,  {Springel} V.,  {Bauer} A.,  {Mocz} P.,  {Munoz} D.~J.,
  {Ohlmann} S.~T.,  {Schaal} K.,   {Zhu} C.,  2016, \mn@doi [\mnras]
  {10.1093/mnras/stv2380}, \href
  {https://ui.adsabs.harvard.edu/abs/2016MNRAS.455.1134P} {455, 1134}

\bibitem[\protect\citeauthoryear{{Papaloizou}, {Nelson}  \&
  {Masset}}{{Papaloizou} et~al.}{2001}]{Papaloizou2001}
{Papaloizou} J.~C.~B.,  {Nelson} R.~P.,   {Masset} F.,  2001, \mn@doi [\aap]
  {10.1051/0004-6361:20000011}, \href
  {https://ui.adsabs.harvard.edu/abs/2001A&A...366..263P} {366, 263}

\bibitem[\protect\citeauthoryear{{Phinney}}{{Phinney}}{2001}]{Phinney2001}
{Phinney} E.~S.,  2001, arXiv e-prints, \href
  {https://ui.adsabs.harvard.edu/abs/2001astro.ph..8028P} {pp
  astro--ph/0108028}

\bibitem[\protect\citeauthoryear{Rajagopal \& Romani}{Rajagopal \&
  Romani}{1995}]{Rajagopal1995}
Rajagopal M.,  Romani R.~W.,  1995, \mn@doi [The Astrophysical Journal]
  {10.1086/175813}, 446, 543

\bibitem[\protect\citeauthoryear{Roedig, Dotti, Sesana, Cuadra  \&
  Colpi}{Roedig et~al.}{2011}]{Roedig2011}
Roedig C.,  Dotti M.,  Sesana A.,  Cuadra J.,   Colpi M.,  2011, \mn@doi
  [Monthly Notices of the Royal Astronomical Society]
  {10.1111/j.1365-2966.2011.18927.x}, 415, 3033

\bibitem[\protect\citeauthoryear{Shi, Krolik, Lubow  \& Hawley}{Shi
  et~al.}{2012}]{Shi2012}
Shi J.-M.,  Krolik J.~H.,  Lubow S.~H.,   Hawley J.~F.,  2012, \mn@doi [The
  Astrophysical Journal] {10.1088/0004-637x/749/2/118}, 749, 118

\bibitem[\protect\citeauthoryear{{Siwek}, {Kelley}  \& {Hernquist}}{{Siwek}
  et~al.}{2020}]{Siwek2020}
{Siwek} M.~S.,  {Kelley} L.~Z.,   {Hernquist} L.,  2020, \mn@doi [\mnras]
  {10.1093/mnras/staa2361}, \href
  {https://ui.adsabs.harvard.edu/abs/2020MNRAS.498..537S} {498, 537}

\bibitem[\protect\citeauthoryear{Springel}{Springel}{2010}]{Springel2010}
Springel V.,  2010, \mn@doi [Monthly Notices of the Royal Astronomical Society]
  {10.1111/j.1365-2966.2009.15715.x}, 401, 791

\bibitem[\protect\citeauthoryear{{Springel}, {Yoshida}  \& {White}}{{Springel}
  et~al.}{2001}]{Springel2001}
{Springel} V.,  {Yoshida} N.,   {White} S. D.~M.,  2001, \mn@doi [\na]
  {10.1016/S1384-1076(01)00042-2}, \href
  {https://ui.adsabs.harvard.edu/abs/2001NewA....6...79S} {6, 79}

\bibitem[\protect\citeauthoryear{{Thun}, {Kley}  \& {Picogna}}{{Thun}
  et~al.}{2017}]{Thun2017}
{Thun} D.,  {Kley} W.,   {Picogna} G.,  2017, \mn@doi [\aap]
  {10.1051/0004-6361/201730666}, \href
  {https://ui.adsabs.harvard.edu/abs/2017A&A...604A.102T} {604, A102}

\bibitem[\protect\citeauthoryear{{Tiede}, {Zrake}, {MacFadyen}  \&
  {Haiman}}{{Tiede} et~al.}{2020}]{Tiede2020}
{Tiede} C.,  {Zrake} J.,  {MacFadyen} A.,   {Haiman} Z.,  2020, arXiv e-prints,
  \href {https://ui.adsabs.harvard.edu/abs/2020arXiv200509555T} {p.
  arXiv:2005.09555}

\bibitem[\protect\citeauthoryear{{Virtanen} et~al.,}{{Virtanen}
  et~al.}{2020}]{2020SciPy-NMeth}
{Virtanen} P.,  et~al., 2020, \mn@doi [Nature Methods]
  {https://doi.org/10.1038/s41592-019-0686-2}, \href {https://rdcu.be/b08Wh}
  {17, 261}

\bibitem[\protect\citeauthoryear{Waskom et~al.,}{Waskom et~al.}{2017}]{seaborn}
Waskom M.,  et~al., 2017, Mwaskom/Seaborn: V0.8.1 (September 2017),
  \mn@doi{10.5281/ZENODO.883859}, \url {https://zenodo.org/record/883859}

\bibitem[\protect\citeauthoryear{Wyithe \& Loeb}{Wyithe \&
  Loeb}{2003}]{Wyithe2003}
Wyithe J. S.~B.,  Loeb A.,  2003, \mn@doi [The Astrophysical Journal]
  {10.1086/375187}, 590, 691

\bibitem[\protect\citeauthoryear{Yu \& Tremaine}{Yu \& Tremaine}{2002}]{Yu2002}
Yu Q.,  Tremaine S.,  2002, \mn@doi [Monthly Notices of the Royal Astronomical
  Society] {10.1046/j.1365-8711.2002.05532.x}, 335, 965

\bibitem[\protect\citeauthoryear{Zrake, Tiede, MacFadyen  \& Haiman}{Zrake
  et~al.}{2021}]{Zrake2021}
Zrake J.,  Tiede C.,  MacFadyen A.,   Haiman Z.,  2021, ]
  {10.3847/2041-8213/abdd1c}, 909, L13

\bibitem[\protect\citeauthoryear{van~der Walt, Colbert  \& Varoquaux}{van~der
  Walt et~al.}{2011}]{vanderWalt2011}
van~der Walt S.,  Colbert S.~C.,   Varoquaux G.,  2011, \mn@doi [Computing in
  Science {\&} Engineering] {10.1109/mcse.2011.37}, 13, 22

\makeatother
\end{thebibliography}

\appendix
\section{Viscosity}
\label{sec:appendix_viscosity}
We model all accretion disks that form in our simulation as alpha disks. Specifically, we use

\begin{equation}
\label{eqn:alpha_visc}
\nu (R)= \alpha h c_s (R) R
\end{equation}
where $\alpha = 0.1$ is a constant, $h$ is the disk scale height, which we vary, and $c_s$ is the locally isothermal sound speed as follows,

\begin{equation}
c_s(R) = h \times \sqrt{\lvert \Phi_{\rm binary} \rvert}.
\end{equation}

The binary potential is defined as follows

\begin{equation}
\Phi_{\rm binary} = -\frac{G M_1}{\lvert \textbf{R} - \textbf{R}_1 \rvert} -\frac{G M_2}{\lvert \textbf{R} - \textbf{R}_2\rvert} .
\end{equation}

The remaining variable in equation (\ref{eqn:alpha_visc}) is $R$. At large radii in the CBD, this should be the distance to the center of the grid, where the gravitational influence of the binary can be reduced to a single point mass at the center of mass of the binary. However, within the cavity region of the disk, and close to any particular sink particle, $R$ should be the distance between the gas particle for which the viscosity is calculated, and its nearest sink particle. This ensures that each CSD is described as an $\alpha$-disk with the same scale height $h$ as the CBD. To achieve this, we define a viscosity $\nu_i$ for each sink particle, and describe the overall viscosity $\nu$ with the expression,

\begin{equation}
\label{eqn:combined_visc}
\frac{1}{\nu^n} = \frac{1}{\nu_1^n} + \frac{1}{\nu_2^n}.
\end{equation}
We choose $n = 5$ and write the viscosities contributed by the individual sink particles, 
$\nu_1 = \alpha h c_s(R_1) r_1$
and
$\nu_2 = \alpha h c_s(R_2) r_2$
where $R_1$ and $R_2$ are the distances between the gas cell where we want to calculate the viscosity and each sink particle. The expression in equation (\ref{eqn:combined_visc}) can then be simplified to,

\begin{equation}
\label{eqn:final_visc}
\nu = \alpha c_s(R_1) h \frac{R_1 R_2}{(R_1^n + R_2^n)^{1/n}}.
\end{equation}
For large enough $n$ (we choose $n=5$), this expression satisfies the requirement that
$\nu \rightarrow \alpha c_s h R$ when $R_1 \simeq R_2$ at large radii, and $ \nu \rightarrow \alpha c_s h R_1$ when $R_1 \ll R_2$.
\label{lastpage}
\end{document}